\documentclass[10pt]{article}
\usepackage{amsmath, amsthm, amssymb }
\usepackage{float,epsfig}
\usepackage{color}
\usepackage{multirow}
\usepackage{cleveref}
\usepackage{braket}
\usepackage{subfigure}
\usepackage[utf8]{inputenc}

\begin{document}

\newtheorem{theorem}{\bf Theorem}[section]
\newtheorem{proposition}[theorem]{\bf Proposition}
\newtheorem{definition}[theorem]{\bf Definition}
\newtheorem{corollary}[theorem]{\bf Corollary}
\newtheorem{example}[theorem]{\bf Example}
\newtheorem{remark}[theorem]{\bf Remark}
\newtheorem{lemma}[theorem]{\bf Lemma}
\newcommand{\nrm}[1]{|\!|\!| {#1} |\!|\!|}

\newcommand{\ba}{\begin{array}}
\newcommand{\ea}{\end{array}}
\newcommand{\dm}[1]{ {\displaystyle{#1} } }
\newcommand{\be}{\begin{equation}}
\newcommand{\ee}{\end{equation}}
\newcommand{\beano}{\begin{eqnarray*}}
\newcommand{\eeano}{\end{eqnarray*}}
\newcommand{\inp}[2]{\langle {#1} ,\,{#2} \rangle}
\def\bmatrix#1{\left[ \begin{matrix} #1 \end{matrix} \right]}
\def \noin{\noindent}
\newcommand{\evenindex}{\Pi_e}

\newcommand{\tb}[1]{\textcolor{blue}{ #1}}
\newcommand{\tm}[1]{\textcolor{magenta}{ #1}}
\newcommand{\tre}[1]{\textcolor{red}{ #1}}



\def \B{{\mathcal B}}
\newcommand{\C}{{\mathbb C}}
\def \H{{\mathcal H}}
\def \L{{\mathsf L}}
\def \O{{\mathcal O}}
\def \P{{\mathcal P}}
\def \Q{{\mathbb Q}}
\newcommand{\R}{{\mathbb R}}
\newcommand{\U}{{\mathrm U}}
\def \X{{\mathcal X}}
\def \Y{{\mathcal Y}}
\def \Z{{\mathcal Z}}

\def \PO{{\mathcal {PO}}}
\def \pf{{\bf Proof: }}
\def \lam{{\lambda}}

\title{Localization of two dimensional quantum walks defined by generalized Grover coins}
\author{ Amrita Mandal\thanks{Department of Mathematics, IIT Kharagpur,Email: mandalamrita55@gmail.com}, \, \, Rohit Sarma Sarkar\thanks{Department of Mathematics, IIT Kharagpur,Email: rohit15sarkar@yahoo.com}, \, \, Bibhas Adhikari\thanks{Corresponding author, Department of Mathematics, IIT Kharagpur,Email: bibhas@maths.iitkgp.ac.in}}
\date{}
\maketitle

\thispagestyle{empty}

\noindent{\bf Abstract.} Localization phenomena of quantum walks makes the propagation dynamics of a walker strikingly different from that corresponding to classical random walks. In this paper, we study the localization phenomena of four-state discrete-time quantum walks on two-dimensional lattices with coin operators as one-parameter orthogonal matrices that are also permutative, a combinatorial structure of the Grover matrix. We show that the proposed walks localize at its initial position for canonical initial coin states when the coin belongs to classes which contain the Grover matrix that we consider in this paper, however, the localization phenomena depends on the coin parameter when the class of parametric coins  does not contain the Grover matrix. \\





\section{Introduction}

In this paper, we study the localization phenomena of discrete-time four-state quantum walks on two dimensional lattices with parametric coin operators as orthogonal permutative matrices (OPMs) of order $4,$ which we call generalized Grover coins. An orthogonal permutative matrix is an orthogonal matrix whose any row is a permutation of any other row, a property of the well-known Grover matrix $G=\frac{1}{2}\left(\boldsymbol{1}_4\boldsymbol{1}_4^\dagger - I_4\right)$ where $\boldsymbol{1}_4$ denotes the all-one (column) vector of dimension $4$ and $I_4$ is the identity matrix of order $4.$ We call the proposed walks as generalized Grover walks. The set of all OPMs of order $4$ is denoted as $\mathcal{O}\mathcal{P}_4$. 

A discrete-time quantum walk (DTQW) is governed by the repeated application of a unitary operator $\U=S_f(C\otimes I)$ to the initial state of the walker, where $S_f$ is called the shift operator, $C$ is called the coin operator, and $\otimes$ denotes the Kronecker (also known as tensor) product of matrices. Thus the evolution operator $\U$ of a proposed walk  acts on the Hilbert space, which is the tensor product of the position space spanned by the quantum states localized at the vertices of the lattice and the coin space whose dimension $4$ gives the internal degree of freedom of the quantum coin. A coined walk is called the Grover walk if coin operator is the Grover matrix. In this paper, we particularly focus on the proposed walks when the coin operator $C\in X_\theta \cup Y_\theta \cup Z_\theta\subset \mathcal{O}\mathcal{P}_4,$ where \be \label{coin:p34_x1} X_\theta =\left\{\bmatrix{\frac{1}{2}\sin\theta&-\frac{1}{2}\sin\theta&\frac{1}{2}(1+\cos\theta)&\frac{1}{2}(1-\cos\theta)\\-\frac{1}{2}\sin\theta&\frac{1}{2}\sin\theta&\frac{1}{2}(1-\cos\theta)&\frac{1}{2}(1+\cos\theta)\\\frac{1}{2}(1-\cos\theta)&\frac{1}{2}(1+\cos\theta)&\frac{1}{2}\sin\theta&-\frac{1}{2}\sin\theta\\\frac{1}{2}(1+\cos\theta)&\frac{1}{2}(1-\cos\theta)&-\frac{1}{2}\sin\theta&\frac{1}{2}\sin\theta}:\theta \in [-\pi,\pi]\right\},\ee 
\be \label{coin:p24_y1} Y_\theta =\left\{ \bmatrix{\frac{1}{2}\sin\theta &\frac{1}{2}(1+\cos\theta)&-\frac{1}{2}\sin\theta&\frac{1}{2}(1-\cos\theta)\\\frac{1}{2}(1-\cos\theta)&\frac{1}{2}\sin\theta&\frac{1}{2}(1+\cos\theta)&-\frac{1}{2}\sin\theta\\-\frac{1}{2}\sin\theta&\frac{1}{2}(1-\cos\theta)&\frac{1}{2}\sin\theta&\frac{1}{2}(1+\cos\theta)\\\frac{1}{2}(1+\cos\theta)&-\frac{1}{2}\sin\theta&\frac{1}{2}(1-\cos\theta)&\frac{1}{2}\sin\theta}:\theta \in [-\pi,\pi] \right\},\ee 
\be  \label{coin:p23_z1} Z_\theta =\left\{\bmatrix{\frac{1}{2}\sin\theta&\frac{1}{2}(1+\cos\theta)&\frac{1}{2}(1-\cos\theta)&-\frac{1}{2}\sin\theta\\\frac{1}{2}(1-\cos\theta)&\frac{1}{2}\sin\theta&-\frac{1}{2}\sin\theta&\frac{1}{2}(1+\cos\theta)\\\frac{1}{2}(1+\cos\theta)&-\frac{1}{2}\sin\theta&\frac{1}{2}\sin\theta&\frac{1}{2}(1-\cos\theta)\\-\frac{1}{2}\sin\theta&\frac{1}{2}(1-\cos\theta)&\frac{1}{2}(1+\cos\theta)&\frac{1}{2}\sin\theta}:\theta \in [-\pi,\pi]\right\},
\ee and the conditional shift operator moves the walker to an adjacent position based on its coin state (see Section \ref{Sec:2}). We emphasize that the parametric matrices in $X_\theta, Y_\theta,$ and $Z_\theta$ can be treated as continuous deformations of the Grover matrix of order $4$ since $C=G$ for $\theta=-\pi/2$ \cite{perpaper22}.

In a few occasions in the literature the Grover walk is generalized by considering coin operators as parametric unitary matrices, which include the Grover matrix as a special case for some particular values of the parameters \cite{watabe2008limit,machida15b,machidachandra15}. In \cite{vstefavnak2012continuous}, the parametric coin operators for three-state quantum walks are proposed by deforming the eigenvalues and the eigenvectors of the Grover matrix. Recently we have proposed to generalize three-state Grover walk on one-dimensional lattice and on cycle graphs by considering orthogonal permutative matrices of order $3$ as parametric coins in \cite{sarkar2020,mandal2022limit}.  The only existing parametric coin operators of order $4$ in the literature is first reported in \cite{inui2004}, and limit distribution and localization of the quantum walks on two-dimensional lattices defined by these parametric coins are studied in \cite{watabe2008limit}. However, our proposal of parametric OPMs as coin operators is significantly different from existing parametric coin operators in literature \cite{inui2004}. The paramertric coins that we consider in this paper preserve the combinatorial structure of the Grover matrix i.e. the permutative structure of the Grover coin and the proposed coins also can be expressed as linear combinations of permutation matrices (see \cite{perpaper22}) which have suitable quantum circuit representation \cite{bataille2022quantum}. 


It is pertinent to investigate whether the characteristics of the Grover walk extend to generalized Grover walks. The localization phenomena of the walker of a quantum walk at a given position is concerned with finding the walker  at that position with a nonzero probability even if the number of walking time-steps tends to infinity \cite{konno2009localization,venegas2012quantum,machida15}. In this paper, we analyze whether the localization property of the Grover walk \cite{tregenna2003controlling}  extends to the proposed generalized Grover walks. We also investigate how does the probability of finding the walker at a given position depend on the values of the coin parameter for the proposed quantum walks. A brief discussion on localization property is given in Section \ref{Sec:2}. 


We show that the proposed walks on infinite lattice exhibit the localization phenomena at the initial position for canonical initial coin states when the coin operator $C\in X_\theta\cup Y_\theta \cup Z_\theta$. Thus we show that the walker can be found at the initial position, which is considered as the vertex $(0,0)$ of the infinite lattice, with a nonzero total time-averaged probability (defined in Section \ref{Sec:2}). This phenomena is also called trapping of the walker \cite{wojcik2012trapping,kollar2015strongly,kollar2020complete}. Next we demonstrate how does this probability value varies with the value of the coin parameter $\theta\in [-\pi, \, \pi]$ for a fixed choice of the initial state. We observe that the maximum and minimum probability values are attained at $\theta =0$ or $\theta = \pm \pi$ when $C\in Y_\theta\cup Z_\theta,$ i.e. coins are permutation matrices and if $C\in X_\theta,$ the maximum and minimum values of the probability are attained at $\theta=\pm \pi/2$ or $0$ or $\pm \pi.$ i.e. coins are $G,P_{(12)(34)}G$ or permutation matrices. Here $P_\sigma$ denotes the permutation matrix associated with the permutation $\sigma\in S_4,$ the symmetric group of order $4.$ Indeed, the $ij$ entry of $P_\sigma$ is $1$ if $\sigma(i)=j,$ otherwise it is $0.$ Later we derive all initial coin states for which the underlying walks with the coins from $ Y_\theta \cup Z_\theta$ have a zero time-averaged probability at the initial position i.e. the origin of the infinite lattice, and hence the walks do not exhibit localization for those initial coin states in contrast to our proof that a walk localizes for any initial coin state when the coin operator $C$ is chosen from $X_\theta$ and $C\neq G.$
We also observe that even though the coins in $ Y_\theta,Z_\theta$ are permutation scaling of coins in $X_\theta$ for any fixed value of $\theta,$ to be precise, $Y_\theta=\{P_{(23)}A_{\theta}P_{(23)} : A_\theta\in X_\theta\}$ and $Z_\theta=\{P_{(13)}A_{\theta}P_{(13)} : A_\theta\in X_\theta\}$, the eigenpairs of the corresponding evolution operators are significantly different. Consequently, the probability of finding the walker varies with the choice of the coin from $X_\theta, Y_\theta, Z_\theta$ for a fixed value of $\theta.$ 

Finally, we show that localization property of the Grover walk corresponding to the canonical initial coin states is not invariant under the permutative property of the coin by finding a set of parametric coin matrices $W_\theta\subset \mathcal{O}\mathcal{P}_4$ given by 
 \be  \label{coin:x3} W_{\theta}=\left\{\bmatrix{\frac{1}{2}(1+\cos \theta)&\frac{1}{2}(1-\cos \theta)&\frac{1}{2}\sin \theta&-\frac{1}{2}\sin \theta\\\frac{1}{2}(1-\cos \theta)&\frac{1}{2}(1+\cos \theta)&-\frac{1}{2}\sin \theta&\frac{1}{2}\sin \theta\\\frac{1}{2}\sin \theta&-\frac{1}{2}\sin \theta&\frac{1}{2}(1-\cos \theta)&\frac{1}{2}(1+\cos \theta)\\-\frac{1}{2}\sin \theta&\frac{1}{2}\sin \theta&\frac{1}{2}(1+\cos \theta)&\frac{1}{2}(1-\cos \theta)}:-\pi\leq \theta\leq \pi\right\}\ee for which quantum walks with canonical initial coin states do not exhibit localization for some coins belonging to $W_\theta.$ Note that $G\not\in W_\theta $ for any $\theta\in [-\pi,\pi]$ and hence the matrices in $W_\theta$ need not be considered as continuous deformation of the Grover coin. 

The organization of the remainder of the paper is as follows. In Section \ref{Sec:2} we briefly discuss the different notions of the localization property as considered  in literature. Then we justify the choice for the definition of localization which we consider in this paper.  In Section \ref{Sec:3} we study the localization property of the proposed quantum walks when the coin operator belongs to $X_\theta, Y_\theta, Z_\theta$ and $W_\theta.$ Then we conclude the paper.


\section{Preliminaries} \label{Sec:2} 
In this section, we discuss the mathematical machinery for analyzing the phenomena of localization of the proposed walks. We first present a brief review of four-state DTQWs on two-dimensional lattices following \cite{inui2004}. Then, we elaborate the notion of localization property of DTQWs developed in several other articles in the literature.

Let \begin{equation}
    Z_N=\{ (x, y)\in \mathbb{Z}^2 : -(N-1)/2\leq x\leq (N-1)/2, -(N-1)/2\leq y\leq (N-1)/2\},
\end{equation} denote the square lattice with $N^2$ vertices, where $N$ is odd. Let $\H_p$ and $\H_c$ denote the $N^2$-dimensional position space and four-dimensional coin space respectively. Then the  proposed walk is defined by the states of the walker as $\ket{\psi(t)}=\U\ket{\psi(t-1)}=\U^{t}\ket{\psi(0)}, t\geq 1$  where the evolution operator is given by $\U=S_f(C\otimes I)$ with initial state $\ket{\psi(0)},$  $S_f=\sum_{(x,y)\in Z_N} S_{x,y}$ where \beano
S_{x,y} &=& \ket{R}\bra{R}\otimes \ket{x-1 (\mbox{mod} N), y} \bra{x,y}  +\ket{L}\bra{L}\otimes \ket{x+1 (\mbox{mod} N), y} \bra{x,y} \\ && +\ket{U}\bra{U}\otimes \ket{x, y-1 (\mbox{mod} N)}\bra{x,y}  +\ket{D}\bra{D}\otimes \ket{x, y+1 (\mbox{mod} N)}\bra{x,y},
\eeano represents the conditional shift operator and $C$ denotes the coin operator. Besides, $\ket{R}=\ket{1}, \ket{L}=\ket{2}, \ket{U}=\ket{3}, \ket{D}=\ket{4}$ denote the chirality of the walker: right, left, up and down respectively, where $\{\ket{1},\ket{2},\ket{3},\ket{4}\}$ denotes the canonical orthonormal ordered basis of the Hilbert space $\H_c$. Then corresponding to the orthonormal basis $\{\ket{S,x,y} : S\in\{R,L,U,D\}, (x,y)\in Z_N\}$ of the total state space, the state of the walker at time $t$ can be written as $$\ket{\psi(t)}=\sum_{S \in \{R,L,U,D\}}\sum_{(x,y)\in Z_N} \alpha_{\ket{S,x,y;t}}\ket{S,x,y}.$$ Then it may be noted that the basis element $\ket{S,x,y}=\ket{S}\otimes \ket{(x,y)}$ can be identified with a canonical basis element $\ket{j},$  $1\leq j\leq 4N^2$ of $4N^2$-dimensional Hilbert space $\C^{4N^2}$ by  $j=4Ny+4x+l(S)+2N^2-2$ and $l(S)=1,2,3,4$ for $S=R,L,U,D$ respectively, see \cite{inui2004}.

Thus the probability of finding the walker at a vertex $(x,y)$ at time $t$ is given by 
\begin{equation}\label{eqn:prob}
    P_t((x,y);\psi(0))= \|\ket{\psi_{(x,y)}(t)}\|_2^2=\sum_{S \in \{R,L,U,D\}}|\alpha_{\ket{S,x,y;t}}|^2
\end{equation} where the initial state $\ket{\psi(0)}$ is known and \begin{eqnarray}\label{psixy} \ket{\psi_{(x,y)}(t)} = \sum_{S \in \{R,L,U,D\}}\alpha_{\ket{S,x,y;t}}\ket{S} 
\end{eqnarray} corresponds to the vector $\bmatrix{\alpha_{\ket{R,x,y;t}}, \alpha_{\ket{L,x,y;t}}, \alpha_{\ket{U,x,y;t}}, \alpha_{\ket{D,x,y;t}}}^T$ of $\C^4.$  Then following \cite{inui2004}, we define localization of the walk at a vertex $(x,y)\in Z_N$ with the help of time-averaged probability \begin{equation}\label{def:aveprob}
    \overline{P}_t((x,y);\psi(0))= \frac{1}{T}\sum_{t=0}^{T-1}  P_t((x,y);\psi(0)), T\geq 1
    \end{equation} as follows. 

\begin{definition}\label{def:loc}
The proposed walk localizes at a vertex $(x,y)\in Z_N$ for some initial state $\ket{\psi(0)}$ if the \textit{total time-averaged probability} $$\overline{P}_N((x,y);\psi(0)) = \lim_{T\rightarrow \infty} \overline{P}_t((x,y);\psi(0)) >0.$$

For infinite lattice, the walk localizes at a vertex $(x,y)$  if \textit{total time-averaged probability}  $$\overline{P}_{\infty}((x,y);\psi(0))=\lim_{N\rightarrow \infty} \overline{P}_N((x,y);\psi(0)) > 0.$$
\end{definition}

The operational meaning of total time-averaged probability is that it captures the proportion of time which the walker ``spends” in any given node $(x,y)$ of the lattice for any initial state $\ket{\psi(0)}.$ We emphasize that localization is defined in the literature in several other ways \cite{higuchi2014spectral, segawa2016generator, tate2019eigenvalues}. For instance, localization is defined in \cite{tate2019eigenvalues} if $\limsup_{t\rightarrow \infty} P_t((x,y);\psi(0))>0$ for quantum walks with periodic evolution operator on  infinite lattice and it can be determined  using the eigenvector projection operators. However, computing eigenfunctions of operators defined on infinite dimensional spaces is challenging. On the other hand, it is easy to verify that if the walk localizes according to Definition \ref{def:loc} then it also localizes according to the definition of localization used in \cite{tate2019eigenvalues}. 

Further, note that $$\overline{P}_{N}((x,y);\psi_S(0))=\sum_{S'\in\{R,L,U,D\}}\overline{P}_{N}(S',(x,y);\psi_S(0)),$$ where  \begin{equation}\label{def:prob_inf}\overline{P}_{N}(S',(x,y);\psi_S(0)) =\lim_{T\rightarrow \infty} \frac{1}{T} \sum_{t=0}^{T-1}  |\alpha_{\ket{S',x,y;t}}|^2\end{equation} denotes the time-averaged probability for the walker to be found at $(x,y)$ with a given coin state $\ket{S'}$ and $\ket{\psi_S(0)}$ denotes the initial state of the walker with $\ket{S}$ as the initial coin state. Then the walk is localized if $\overline{P}_N(S',(x,y);\psi_S(0)) >0$ for some $S'.$ Obviously, equation (\ref{def:prob_inf}) is valid for infinite lattice by considering $N\rightarrow \infty.$

The Grover walk, when the coin operator is the Grover matrix, is well studied in literature \cite{inui2004, komatsu2019eigenvalues}. It is shown that localization of Grover walk on two-dimensional infinite lattice depends on initial state of the walk and it is also speculated that the asymptotic behavior of the walk depends on the eigenvalues of the evolution operator. Then, based on the degeneracy of eigenvalues of the evolution operator, a necessary and sufficient condition is obtained in \cite{tate2019eigenvalues} for quantum walks on infinite lattices corresponding to periodic evolution operator that does not localize at any vertex.

It is found that the eigenvalues of the evolution matrix $\U$ for the Grover walk can be derived from the eigenvalues of another matrix $\U_{n,m}=D_{n,m}G$ using Fourier transform, where $D_{n,m}$ is a unitary diagonal matrix and $G$ is the Grover matrix \cite{inui2004}. The same holds true when the Grover matrix is replaced by generalized Grover coins as described in the following proposition. The proof is similar to the case when the coin operator is of order $3$ as considered in $\cite{sarkar2020}$, hence we omit the proof.

\begin{proposition} \label{prop:eig vec U}
The two dimensional quantum walk operator $\mathrm{U}=S_f(C\otimes I)$ has eigenvalues $\lambda_{n,m,j}$ with a corresponding eigenvector $\ket{\eta_{n,m,j}}=\ket{v_{n,m,j}} \otimes \ket{\phi_{n}, \phi'_{m}}$ where $\lambda_{n,m,j}$ is an eigenvalue of $\U_{n,m}=D_{n,m}C$ corresponding to an eigenvector $\ket{v_{n,m,j}}$, $\ket{\phi_{n}}=\sum_{x=-\frac{N-1}{2}}^{\frac{N-1}{2}}e^{-ikx}\ket{x}$,and $\ket{\phi'_{m}}=\sum_{y=-\frac{N-1}{2}}^{\frac{N-1}{2}}e^{-ik'y}\ket{y}; k=\frac{2\pi n}{N}$,$k'=\frac{2\pi m}{N}$, $n,m \in \{0,1,...,N-1\}$, and $D_{n,m}=\mbox{diag}(\omega^{-n},\omega^{n},\omega^{-m},\omega^{m})$  where $\omega = e^{\frac{2\pi i}{N}}$ for $i=\sqrt{-1}.$ 
\end{proposition}

Then the following corollary describes each entry of eigenvectors of ${\U}$ from the eigenvectors of ${\U}_{n,m}.$ 

\begin{corollary} \label{cor:eigvec U}
Let $(\lam_{n,m,k}, \ket{\eta_{n,m,k}}),$ $k\in\{1,2,3,4\},$ $n,m\in\{0,1,\hdots, N-1\}$ denote eigenpairs of $\U.$ If $\ket{\eta_{n,m,k}}=[\eta_{r,n,m,k}],$ $r=1,\hdots, 4N^2$ then 
$$\eta_{r,n,m,k}=\frac{v_{s,n,m,k}\omega^{-(nx+my)}}{N{\|\ket{v_{n,m,k}}\|_2}},$$ where $(x,y)\in Z_N$ and $s\in\{1,2,3,4\}$  satisfy $r=4Ny+4x+s+2N^2-2$ and $\ket{v_{n,m,k}}=[v_{s,n,m,k}]$ is an eigenvector of $\U_{n,m}$ associated with the eigenvalue $\lam_{n,m,k}.$ \end{corollary} 

It is obvious to check that  the eigenvectors of $\U$ as described in Proposition \ref{prop:eig vec U} are orthonormal.


\section{Localization of two dimensional four state quantum walks with generalized Grover coins}\label{Sec:3}

In this section, we investigate localization phenomena of the proposed walks. First, we consider the walks defined by coins from $Y_{\theta},X_{\theta}$, and  $Z_{\theta}$ and then for  coins from $W_{\theta},$ when $\theta\in  [-\pi,\pi].$ We establish that the walks with coins from $Y_{\theta},X_{\theta}, Z_{\theta}$ exhibit the localization property at the initial position $(0,0)\in \mathbb{Z}\times \mathbb{Z}$ for the given canonical initial coin states after providing a computable formula of the total time-averaged probability for each $\theta$. Obviously, total time-averaged probability depends on the eigenpairs of the evolution operator. For finite lattice, the time-averaged probability also depends on the size of the lattice, and hence we focus on the infinite lattice and the total time-averaged probability is computed by approximating it using an integral formula. Indeed, the formula depends on the eigenvectors corresponding to constant eigenvalues and the initial state of the evolution operator (see Remark \ref{remark:end}). Moreover, we characterize those initial states for which the time-averaged probability has zero or nonzero value at the position $(0,0)$. Finally, the influence of the coin operator is observed for a fixed initial state by plotting the total time-averaged probability for several values of the coin parameter, which is one of the prime objectives of the paper.


 \subsection{With coins from $Y_\theta$}

Consider the walks when the coin operator $C\in Y_{\theta},$ $\theta\in [-\pi, \, \pi].$ Note that if $\mathrm{U}=S_f(C\otimes I)$ denotes the evolution operator then  \begin{equation}\label{p24y1}\mathrm{U}_{n,m}=D_{n,m}C=\bmatrix{\frac{1}{2}\sin{\theta}\omega^{-n} & \frac{1+\cos{\theta}}{2}\omega^{-n} & -\frac{1}{2}\sin{\theta}\omega^{-n} & \frac{1-\cos{\theta}}{2}\omega^{-n} \\
\frac{1-\cos{\theta}}{2}\omega^{n} & \frac{1}{2}\sin{\theta}\omega^{n} & \frac{1+\cos{\theta}}{2}\omega^{n} & -\frac{1}{2}\sin{\theta}\omega^{n}\\
-\frac{1}{2}\sin{\theta}\omega^{-m} & \frac{1-\cos{\theta}}{2}\omega^{-m} & \frac{1}{2}\sin{\theta}\omega^{-m} & \frac{1+\cos{\theta}}{2}\omega^{-m}\\
\frac{1+\cos{\theta}}{2}\omega^{m} & -\frac{1}{2}\sin{\theta}\omega^{m} & \frac{1-\cos{\theta}}{2}\omega^{m} & \frac{1}{2}\sin{\theta}\omega^{m}}.\end{equation}  First we derive the eigenvalues of $\mathrm{U}_{n,m},$ $m,n\in \{0, \hdots, N-1\}$ as follows.

\begin{lemma} \label{lemma:eig u_(n,m)}
Consider $\mathrm{U}_{n,m}$ from equation (\ref{p24y1}), where $\theta \neq 0,\pm \pi$. Then a complete set of orthogonal eigenpairs $(\lambda_{n,m,k}, \ket{v_{n,m,k}}), k=1,\hdots,4$ of $\mathrm{U}_{n,m}$ are 
\begin{eqnarray*} && \lambda_{n,m,1}=-1, \lambda_{n,m,2}= 1, 
 \lambda_{n,m,3}=\frac{\sin{\theta}(\cos{\zeta_n}+\cos{\zeta_m})-i\sqrt{4-\sin^2{\theta}(\cos{\zeta_n}+\cos{\zeta_m})^2}}{2}, \\
&& \lambda_{n,m,4} = \frac{\sin{\theta}(\cos{\zeta_n}+\cos{\zeta_m})+i\sqrt{4-\sin^2{\theta}(\cos{\zeta_n}+\cos{\zeta_m})^2}}{2},
\end{eqnarray*}
 $$\ket{v_{n,0,k}}= \begin{cases} 
\bmatrix{\dfrac{(1+\cos{\theta}+\sin{\theta})\omega^{-n}}{(1+\cos{\theta})\omega^{-n}+\sin{\theta}}, & -\dfrac{(1+\cos{\theta}+\sin{\theta})}{(1+\cos{\theta})\omega^{-n}+\sin{\theta}}, &
1, & -1
}^T \,\, \mbox{if} \,\,  k=1 \\
\bmatrix{\dfrac{(1+\cos{\theta}-\sin{\theta})\omega^{-n}}{(1+\cos{\theta})\omega^{-n}-\sin{\theta}}, & \dfrac{(1+\cos{\theta}-\sin{\theta})}{(1+\cos{\theta})\omega^{-n}-\sin{\theta}}, &
1, & 1
}^T \,\, \mbox{if} \,\, k=2 \\
\left[\left(\dfrac{(1-\cos{\theta})\lambda_{n,0,k}\omega^{-n}-\sin{\theta}}{1-\cos{\theta}-\sin{\theta}\lambda_{n,0,k}\omega^n}\right)\left(\dfrac{(1+\cos{\theta})-\lambda_{n,0,k}\sin{\theta}}{(1+\cos{\theta})\lambda_{n,0,k}\omega^{-n}-\sin{\theta}}\right)\right., \\
\hfill{\dfrac{(1+\cos{\theta})-\lambda_{n,0,k}\sin{\theta}}{(1+\cos{\theta})\lambda_{n,0,k}\omega^{-n}-\sin{\theta}}, 1,} \\
\hfill{\dfrac{2}{(1-\cos{\theta}-\lambda_{n,0,k}\sin{\theta}\omega^n)}\left(\dfrac{(1+\cos{\theta}-\lambda_{n,0,k}\sin{\theta})(\lambda_{n,0,k}^2-\lambda_{n,0,k}\sin{\theta}\cos{\zeta_n})}{(1+\cos{\theta})\lambda_{n,0,k}\omega^{-n}-\sin{\theta}}\right.}\\
\hfill{\left. \left. +\dfrac{\sin{\theta}-(1+\cos{\theta})\lambda_{n,0,k}\omega^n}{2}\right)\right]^T \,\, \mbox{if} \,\, k=3,4,} \\
\end{cases}$$ if $n > 0,$ and 
\begin{eqnarray*}\ket{ v_{n,m,k}}= &&\left[ \left(\dfrac{(1-\cos{\theta})\lambda_{n,m,k}\omega^{-n}-\sin{\theta}}{(1-\cos{\theta})-\sin{\theta}\lambda_{n,m,k}\omega^n}\right)\left(\dfrac{1+\cos{\theta}-\lambda_{n,m,k}\sin{\theta}\omega^m}{(1+\cos{\theta})\lambda_{n,m,k}\omega^{-n}-\sin{\theta}}\right), \right.\\
&& \hfill{\left.\dfrac{1+\cos{\theta}-\lambda_{n,m,k}\sin{\theta}\omega^m}{(1+\cos{\theta})\lambda_{n,m,k}\omega^{-n}-\sin{\theta}}, 
 1, \,\, \dfrac{(1+\cos{\theta})\lambda_{n,m,k}\omega^{m}-\sin{\theta}}{(1+\cos{\theta})-\sin{\theta}\lambda_{n,m,k}\omega^{-m}} \right]^T}\end{eqnarray*} if $n,m>0$ and $k\in\{1,2,3,4\},$ where $\zeta_q=2\pi q/N,$ $q \in \{m,n\}$.
\end{lemma}

\pf Note that $\omega^q=e^{i\zeta_q}$. 
We also note that the characteristic polynomial of ${\U}_{n,m}$ is given by $$\chi_{{\U}_{n,m}}(\lambda)=\lambda^4-\sin{\theta}(\cos{\zeta_m}+\cos{\zeta_n})\lambda^3+\sin{\theta}(\cos{\zeta_m}+\cos{\zeta_n})\lambda-1.$$ By calculating the roots, we get the eigenvalues and  the corresponding eigenvectors can be obtained by solving the system of equations $(\mathrm{U}_{n,m}-\lambda_{n,m,k}I_4)X=0$ for $X.$  Since $\U_{n,m}$ is a unitary matrix, eigenvectors of $\U_{n,m}$ corresponding to different eigenvalues are orthogonal. $\hfill{\square}$

\begin{lemma} \label{lemma:eig u_(n,m) per}
Consider $\mathrm{U}_{n,m}$ from equation (\ref{p24y1}) , where $\theta = 0,\pm \pi$. Then the set of orthogonal eigenpairs $(\lambda_{n,m,k}, \ket{v_{n,m,k}}), k=1,\hdots,4$ of $\mathrm{U}_{n,m}$ are as follows.
For $\theta=0,$
\beano 
& \lambda_{n,m,1}=-1, \lambda_{n,m,2}= 1, \lambda_{n,m,3}=-i,\lambda_{n,m,4} =i,\\
&\ket{v_{n,m,k}}=\bmatrix{\lambda_{n,m,k}\omega^{-m}, \,\,\lambda^2_{n,m,k}\omega^{n-m}, \,\, \lambda^3_{n,m,k}\omega^{-m},\,\, 1}^T,
\eeano
and for $\theta=\pm \pi,$
\beano &\lambda_{n,m,1}=-1, \lambda_{n,m,2}= 1, \lambda_{n,m,3}=-i,\lambda_{n,m,4} =i ,\\
&\ket{v_{n,m,k}}=\bmatrix{\lambda^3_{n,m,k}\omega^{-n}, \,\,\lambda^2_{n,m,k}, \,\, \lambda_{n,m,k}\omega^{-m},\,\, 1}^T,
\eeano
where $\zeta_q=2\pi q/N$ and $\omega^q=e^{i\zeta_q},$ $q \in \{m,n\}.$ 
\end{lemma}
\pf The proof is computational and  easy to verify. $\hfill{\square}$

Then $\ket{\eta_{n,m,k}},$ $1\leq k \leq 4,$ $0\leq m,n \leq N-1$ form a set of orthonormal eigenvectors corresponding to the eigenvalues $\lam_{n,m,k}$ of $\U$, and hence $$\U=\sum_{n,m,k} \lam_{n,m,k} \ket{\eta_{n,m,k}} \bra{\eta_{n,m,k}}.$$ Consequently, 
\beano
\ket{\psi(t)} = \U^t\ket{\psi(0)}&=&\sum_{n,m,k}{\lambda_{n,m,k}^t} \ket{\eta_{n,m,k}}\bra{\eta_{n,m,k}}\ket{\psi(0)}\\
&=&\sum_{j=1}^{4N^2}\sum_{n,m,k}{\lambda_{n,m,k}^t}\overline{\eta}_{j,n,m,k}{\psi_j(0)}\ket{\eta_{n,m,k}} \\
&=& \sum_{j=1}^{4N^2}\sum_{n,m,k} \sum_{S \in \{R,L,U,D\}}\sum_{(x,y)\in Z_N} {\lambda}^t_{n,m,k}\overline{\eta}_{j,n,m,k}\psi_j(0)\eta_{r,n,m,k}\ket{S,x,y},
\eeano
where $r=4Ny+4x+l(S)+2N^2-2.$ Then since the eigenvectors of the proposed evolution operator $\U$ can be obtained by the eigenvectors of $\U_{n,m}$ derived in Lemma \ref{lemma:eig u_(n,m)}, the wave function of the proposed DTWQ can be obtained by employing Corollary \ref{cor:eigvec U}. Indeed,
$$\ket{\psi(t)} = \sum_{S \in \{R,L,U,D\}}\sum_{(x,y)\in Z_N} \alpha_{\ket{S,x,y;t}}\ket{S,x,y}$$ where 
\begin{eqnarray} \label{coeffi:alpha_S}
\alpha_{\ket{S,x,y;t}}
&=& \sum_{j=1}^{4N^2}\sum_{n,m,k}{\lambda}^t_{n,m,k}\overline{\eta}_{j,n,m,k}\psi_j(0)\eta_{r,n,m,k}
\end{eqnarray}
where $j=4Ny'+4x'+l(S')+2N^2-2,$ and  $\psi_j(0)=\alpha_{\ket{S',x',y';0}},$ $(x',y')\in Z_N,S'\in \{R,L,U,D\}.$ Then from equations (\ref{psixy}) and (\ref{coeffi:alpha_S}) we have 
\begin{equation} \label{eqn:vv}
\ket{\psi_{(x,y)}(t)} = \frac{1}{N^2} \sum_{n,m,k} {\lambda}^t_{n,m,k}\frac{\ket{v_{n,m,k}}\bra{v_{n,m,k}}}{\langle v_{n,m,k}\vert v_{n,m,k}\rangle}\ket{\psi_{(0,0)}(0)}\omega^{-(nx+my)},
\end{equation} 
and $P_t((x,y);\psi(0))={\|\ket{\psi_{(x,y)}(t)}\|_2}^2$ follows from equation (\ref{eqn:prob}), which can be computed numerically. An explicit analytical expression of it in terms of the coin parameter $\theta$ is hard to obtain due to the cumbersome expressions of eigenvectors $\ket{v_{n,m,k}}$ of $\U_{n,m}$  given by Lemma \ref{lemma:eig u_(n,m)}.



We now consider calculating the total time-averaged probability for finding the walker at the initial position when the lattice is infinite following \cite{inui2004,inui2005localization}. We introduce some notations that enable us to derive a compact expression of $\overline{P}_t((x,y); \psi(0))$ utilizing the fact that eigenvalues of $\U$ are repeated for different pairs of $(n,m).$ We refer to different coefficients $\alpha_{\ket{S,x,y;t}}$ given in equation (\ref{coeffi:alpha_S}) as follows: 
\begin{equation}\label{eqn:formulac}c_{r,j,0,0,k}= \frac{v_{l(S),0,0,k}\overline{v}_{l(S'),0,0,k}\psi_j(0)}{\langle v_{0,0,k} \vert v_{0,0,k}\rangle}, \,\, \,\, c_{r,j,n,m,k}=\sum_{(n',m') \in \Omega(n,m)} \frac{v_{l(S),n',m',k}\overline{v}_{l(S'),n',m',k}\psi_j(0)}{\langle v_{n',m',k} \vert v_{n',m',k}\rangle}
\end{equation}  where the second expression is defined for $n>0$ and $m>0,$ and $\Omega(n,m)=\{(n',m') : \Lambda(\U_{n,m})=\Lambda(\U_{n',m'})\},$ $\Lambda(X)$ denotes the spectrum of a matrix $X.$ Indeed, $\Omega(n,m)=\{(n,0,)(0,n),(N-n,0),(0,N-n)\}$ if $m=0;$ $\Omega(n,m)= \{(n,n,),(n,N-n), (N-n,n),(N-n,N-n)\}$ if $n=m;$ and $\Omega(n,m) = \{(n,m),(n,N-m),(N-n,m),(N-n,N-m), (m,n),(m,N-n),(N-m,n),(N-m,N-n)\}$ otherwise; follows from the fact that $ \cos{\zeta_{n'}}+\cos{\zeta_{m'}}=\cos{\zeta_n}+\cos{\zeta_m}$ when $(n',m') \in \Omega(n,m),$ $\zeta_q=2\pi q/N, q\in\{n,m\}.$




 
Thus from (\ref{coeffi:alpha_S}) we obtain 
\begin{align} \label{eqn:wave S}
\begin{split}
    \alpha_{\ket{S,x,y;t}}=&\frac{1}{N^2}\sum_{j=1}^{4N^2}\left[C_{r,j,1}(-1)^t+C_{r,j,2} + \sum_{n=1}^{\frac{N-1}{2}}\sum_{k=3}^4 c_{r,j,n,0,k}{\lambda}^t_{n,0,k} \right. \\ 
    &\left.+ \sum_{n=1}^{\frac{N-1}{2}}\sum_{k=3}^4 c_{r,j,n,n,k}{\lambda}^t_{n,n,k}
                 +\sum_{n=1}^{\frac{N-3}{2}}\sum_{m=n+1}^{\frac{N-1}{2}}\sum_{k=3}^4 c_{r,j,n,m,k}{\lambda}^t_{n,m,k}\right],
                 \end{split}
\end{align}
where
\begin{eqnarray}
C_{r,j,1} &=& \sum_{k=1,3,4} c_{r,j,0,0,k} +\sum_{n=1}^{\frac{N-1}{2}} c_{r,j,n,0,1} +\sum_{n=1}^{\frac{N-1}{2}} c_{r,j,n,n,1} +\sum_{n=1}^{\frac{N-3}{2}}\sum_{m=n+1}^{\frac{N-1}{2}} c_{r,j,n,m,1}, \label{wave coeffi:1} \\
C_{r,j,2} &=& c_{r,j,0,0,2}+\sum_{n=1}^{\frac{N-1}{2}} c_{r,j,n,0,2}+\sum_{n=1}^{\frac{N-1}{2}} c_{r,j,n,n,2} +\sum_{n=1}^{\frac{N-3}{2}}\sum_{m=n+1}^{\frac{N-1}{2}} c_{r,j,n,m,2}. \label{wave coeffi:2}
\end{eqnarray}

Now we determine the time-averaged probability for the  walker to be found at the initial position vertex $(0,0)$ with coin state $\ket{S'}$ when the initial coin state of the walker is $\ket{S}$ for the proposed DTQWs,  $S,S'\in\{R,L,U,D\}.$  


First we have the following lemma which is easy to prove.
\begin{lemma}\label{lem:limit}
Let $\lambda_{n,m,k} \neq \overline{\lambda}_{n',m',k'}$ be two eigenvalues of $\U.$ Then  
$$\lim_{T \rightarrow \infty}\sum_{t=0}^{T-1}\frac{(\lambda_{n,m,k})^t(\overline{\lambda}_{n',m',k'})^t}{T}=0.$$
\end{lemma}


Then by Lemma \ref{lem:limit} and equation (\ref{eqn:wave S}), 
we obtain
\begin{eqnarray}\label{eqn:fcase}
\overline{P}_N(S',(0,0);\psi_S(0)) &=& \frac{1}{N^4} \left[\left|\sum_{j=1}^{4N^2} C_{r,j,1}\right|^2+\left|\sum_{j=1}^{4N^2} C_{r,j,2}\right|^2 \right.\nonumber\\
&&\left.+ \sum_{n=1}^{\frac{N-1}{2}}\sum_{k=3}^4 \left|\sum_{j=1}^{4N^2} c_{r,j,n,0,k}\right|^2 + \sum_{n=1}^{\frac{N-1}{2}}\sum_{k=3}^4 \left|\sum_{j=1}^{4N^2} c_{r,j,n,n,k}\right|^2 \right.\nonumber\\
&& \left. +\sum_{n=1}^{\frac{N-3}{2}}\sum_{m=n+1}^{\frac{N-1}{2}}\sum_{k=3}^4\left|\sum_{j=1}^{4N^2} c_{r,j,n,m,k}\right|^2\right] \label{eq:P(s,pi,N)}.
\end{eqnarray}

\begin{remark}{(Dependence of $\overline{P}_\infty(S',(0,0);\psi_S(0))$ on the constant eigenvalues of $\U$)}\label{remark:ev}
From equation (\ref{eq:P(s,pi,N)}) it follows that the $3$rd and $4$th terms in the expression of $\overline{P}_N(S',(0,0);\psi_S(0))$ are of order $N^{-3},$ and the $5$-th term is of order $N^{-2}.$ Thus all these terms vanish when $N \rightarrow \infty$ and hence the probability of observing the walker at the vertex $(0,0),$ the initial position of the walker, depends only on the constant eigenvalues of the evolution operator $\U$ for the infinite lattice.  Although for finite lattices, the time-averaged probability can be computed numerically using equation (\ref{eqn:fcase}), however it depends on the size of the lattice $N^2.$ 
\end{remark}

Thus for infinite lattice, the explicit formula for finding the walker with a coin state $\ket{S'}$ for any initial canonical coin state $\ket{S}$ can be obtained  in terms of the entries of the eigenvectors corresponding to constant eigenvalues of $\U$ as follows.  
 
\begin{eqnarray}\label{def:finalp}
\overline{P}_\infty(S',(0,0);\psi_S(0)) &=& \lim_{N\rightarrow \infty} \overline{P}_N(S',(0,0);\psi_S(0))\\ &=& \lim_{N \rightarrow \infty}\frac{1}{N^4} \left(\left|\sum_{j=1}^{4N^2} C_{r,j,1}\right|^2+\left|\sum_{j=1}^{4N^2} C_{r,j,2}\right|^2\right) \nonumber\\
&=&\lim_{N \rightarrow \infty}\frac{1}{N^4} \left( \left|C_{r,j,1}\right|^2+\left| C_{r,j,2}\right|^2\right)\nonumber,
\end{eqnarray} 
where $ S,S'\in \{R, L, U, D\}$ and $r=l(S')+2N^2-2, j=l(S)+2N^2-2.$

Now we determine time-averaged probability for finding the walker on the infinite lattice at the initial position $(0,0)$ with coin state $\ket{S'}$ when the initial coin state is $\ket{S}$ and coin parameter $\theta \in (-\pi,\pi),\theta\neq 0.$

\begin{theorem}
 $\overline{P}_\infty(S',(0,0);\psi_S(0))=\dfrac{1}{8}$ if $S=S'\in\{R,L,U,D\},$ for $\theta\neq 0,\pm \pi.$
\end{theorem}
\pf 
Let $\ket{S}=\ket{S'}=\ket{R}.$ Note that, if $\ket{\psi_R(0)}=[\psi_j(0)], 1\leq j\leq 4N^2$ then  
$\psi_{2N^2-1}(0)=1$ and $\psi_j(0)=0$ if $j \neq 2N^2-1.$ 
Then by equations (\ref{wave coeffi:1}), (\ref{wave coeffi:2}) and  (\ref{def:finalp}) we have
\beano
\overline{P}_\infty(R,(0,0);\psi_R(0)) &=&\lim_{N \rightarrow \infty}\frac{1}{N^4} \left(\left| C_{r,j,1}|^2+| C_{r,j,2}\right|^2\right)\\ 
 &=&\lim_{N \rightarrow \infty}\frac{1}{N^4}\left(\left|\sum_{n=1}^{(N-3)/2}\sum_{m=n+1}^{(N-1)/2} c_{r,j,n,m,1}\right|^2+\left|\sum_{n=1}^{(N-3)/2}\sum_{m=n+1}^{(N-1)/2} c_{r,j,n,m,2}\right|^2\right).
 \eeano
$r,j=2N^2-1.$ 

Clearly the constant and single summation terms in $C_{r,j,1}$ and  $C_{r,j,2}$ vanishes while $N \rightarrow \infty$ in the above expression. 
Now from Lemma \ref{lemma:eig u_(n,m)} we have
\beano \ket{v_{n,m,1}} &=& \bmatrix{\dfrac{(1+\cos{\theta}+\sin{\theta}\omega^m)\omega^{-n}}{(1+\cos{\theta})\omega^{-n}+\sin{\theta}}, & \,\,-\dfrac{(1+\cos{\theta}+\sin{\theta}\omega^m)}{(1+\cos{\theta})\omega^{-n}+\sin{\theta}}, \,\, 1, &
-\omega^m}^T, \\
\ket{v_{n,m,2}} &=& \bmatrix{\dfrac{(1+\cos{\theta}-\sin{\theta}\omega^m)\omega^{-n}}{(1+\cos{\theta})\omega^{-n}-\sin{\theta}}, & \dfrac{(1+\cos{\theta}-\sin{\theta}\omega^m)}{(1+\cos{\theta})\omega^{-n}-\sin{\theta}}, &
1, &
\omega^m}^T\eeano
for $\theta\neq 0,\pm \pi.$ Thus using equation (\ref{eqn:formulac}) we have 
$c_{r,j,n,m,1}=c_{r,j,n,m,2}=2.$ where $r,j=2N^2-1.$
 Hence 
 \beano \overline{P}_\infty(R,(0,0);\psi_R(0)) =\lim_{N \rightarrow \infty}\frac{2}{N^4}\left|\sum_{n=1}^{(N-3)/2}\sum_{m=n+1}^{(N-1)/2}2 \right|^2
 =2\lim_{N \rightarrow \infty}\frac{1}{N^4}\left(\frac{(N-1)}{2}\frac{(N-3)}{2} \right)^2 =\frac{1}{8}.\eeano Similarly the proof follows for other cases. $\hfill{\square}$

Next we consider finding $\overline{P}_\infty(S',(0,0);\psi_S(0))$ when $S\neq S'.$ Indeed note that it ultimately boils down computing the constants $c_{r,j,n,m,k}$ as follows. From equation (\ref{def:finalp}) we have
\begin{align*}
\overline{P}_\infty(S',(0,0);\psi_S(0))=&\lim_{N\rightarrow \infty}\frac{1}{N^4} \left[\left| C_{r,j,1}\right|^2+\left| C_{r,j,2}\right|^2\right]\\=&\lim_{N \rightarrow \infty}\frac{1}{N^4}\left(\left|\sum_{n=1}^{\frac{N-3}{2}}\sum_{m=n+1}^{\frac{N-1}{2}} c_{r,j,n,m,1}\right|^2+\left|\sum_{n=1}^{\frac{N-3}{2}}\sum_{m=n+1}^{\frac{N-1}{2}} c_{r,j,n,m,2}\right|^2\right),\end{align*} 
where $ S,S'\in \{R, L, U, D\}$ and $r=l(S')+2N^2-2, j=l(S)+2N^2-2.$
Now from equation (\ref{eqn:formulac}) we have
 $$c_{r,j,n,m,k}
=\sum_{(n',m') \in \Omega(n,m)} \frac{v_{l(S'),n',m',k}\overline{v}_{l(S),n',m',k}}{\|\ket{v_{n',m',k}}\|_2^2}, k=1,2$$ whose values can be obtained by placing the values of $v_{l(S),n',m',k}$ and $v_{l(S'),n',m',k}$ from Lemma \ref{lemma:eig u_(n,m)}, where $r=l(S')+2N^2-2$ and $j=l(S)+2N^2-2,1\leq l(S), l(S')\leq 4.$ Indeed for $\theta\neq 0,\pm \pi$,

$$
c_{r,j,n,m,1} = \begin{cases} -\dfrac{2\left(\cos{\zeta_n}+\cos{\zeta_m}+2\sin{\theta}\cos{\zeta_n}\cos{\zeta_m}\right)}{2+\sin{\theta}\left(\cos{\zeta_n}+\cos{\zeta_m}\right)}\,\, \mbox{if}\,\, \{l(S),l(S')\}\in\{\{1,2\},\{3,4\}\},l(S)\neq l(S'),\\
\dfrac{2\left(1+\cos{\theta}+(1-\cos{\theta})\cos{\zeta_n}\cos{\zeta_m}+\sin{\theta}(\cos{\zeta_n}+\cos{\zeta_m})\right)}{2+\sin{\theta}\left(\cos{\zeta_n}+\cos{\zeta_m}\right)} \,\, \mbox{if}\,\, l(S),l(S')\in\{1,3\},l(S)\neq l(S'),\\
\dfrac{-2\left(\cos{\zeta_n}+\cos{\zeta_m}+\sin{\theta}+\sin{\theta}\cos{\zeta_n}\cos{\zeta_m}\right)}{2+\sin{\theta}\left(\cos{\zeta_n}+\cos{\zeta_m}\right)} \,\, \mbox{if}\,\, \{l(S),l(S')\}\in\{\{1,4\},\{2,3\}\},l(S)\neq l(S'),\\
 \dfrac{2\left((1+\cos{\theta})\cos{\zeta_n}\cos{\zeta_m}+(1-\cos{\theta})+\sin{\theta}(\cos{\zeta_n}+\cos{\zeta_m})\right)}{2+\sin{\theta}(\cos{\zeta_n}+\cos{\zeta_m})}\,\, \mbox{if}\,\, l(S),l(S')\in\{2,4\},l(S)\neq l(S'),\\
2 ,\,\, \mbox{if}\,\, l(S),l(S')\in\{1,2,3,4\}, l(S)=l(S'),
\end{cases}$$
$$c_{r,j,n,m,2}= \begin{cases} \dfrac{2\left(\cos{\zeta_n}+\cos{\zeta_m}-2\sin{\theta}\cos{\zeta_n}\cos{\zeta_m}\right)}{2-\sin{\theta}\left(\cos{\zeta_n}+\cos{\zeta_m}\right)}\,\, \mbox{if}\,\, \{l(S),l(S')\}\in\{\{1,2\},\{3
,4\}\},l(S)\neq l(S')\\
\dfrac{2\left(1+\cos{\theta}+(1-\cos{\theta})\cos{\zeta_n}\cos{\zeta_m}-\sin{\theta}(\cos{\zeta_n}+\cos{\zeta_m})\right)}{2-\sin{\theta}\left(\cos{\zeta_n}+\cos{\zeta_m}\right)}\,\, \mbox{if}\,\, l(S),l(S')\in\{1,3\},l(S)\neq l(S')\\
\dfrac{2\left(\cos{\zeta_n}+\cos{\zeta_m}-\sin{\theta}-\sin{\theta}\cos{\zeta_n}\cos{\zeta_m}\right)}{2-\sin{\theta}\left(\cos{\zeta_n}+\cos{\zeta_m}\right)}\,\, \mbox{if}\,\, \{l(S),l(S')\}\in\{\{1,4\},\{2,3\}\},l(S)\neq l(S')\\
  \dfrac{2\left((1+\cos{\theta})\cos{\zeta_n}\cos{\zeta_m}+(1-\cos{\theta})-\sin{\theta}(\cos{\zeta_n}+\cos{\zeta_m})\right)}{2-\sin{\theta}(\cos{\zeta_n}+\cos{\zeta_m})}\,\, \mbox{if}\,\, l(S),l(S')\in\{2,4\},l(S)\neq l(S').\\
2 ,\,\, \mbox{if}\,\,l(S),l(S')\in\{1,2,3,4\}, l(S)=l(S').
\end{cases}$$

Observe that due to complicated expressions of $c_{r,j,n,m,k}, k=1,2$ it is not feasible to come up with a value for $\overline{P}_\infty(S';(0,0),\psi_S(0)))$ when $S\neq S'$ for all $\theta\in (-\pi, \pi), \theta\neq 0.$ However as $N\rightarrow \infty$, the limiting value for the sum of $c_{r,j,n,m,k}$ can be approximated by Riemann integration. For example, consider of $c_{r,j,n,m,1}$ where $l(S),l(S')\in\{1,2\},l(S)\neq l(S').$ Then,
$$\lim_{N \rightarrow \infty}\frac{1}{N^2}\sum_{n=1}^{\frac{N-3}{2}}\sum_{m=n+1}^{\frac{N-1}{2}} c_{r,j,n,m,1}=\frac{1}{8\pi^2}\int_{0}^{\pi}\int_{0}^{\pi}-\frac{2(\cos{x}+\cos{y}+2\sin{\theta}\cos{x}\cos{y})}{2+\sin{\theta}(\cos{x}+\cos{y})}dx \, dy.$$


Then the total time-averaged probability for finding the walker at $(0,0)$ is given by $$\overline{P}_\infty ((0,0);\psi_S(0))=\sum_{S'\in\{R,L,U,D\}} \overline{P}_\infty(S',(0,0);\psi_S(0))$$ when the initial coin state of the walker is $S\in\{R,L,U,D\},$ can be obtained after evaluating the Riemann integration for specific values of $\theta,\theta \neq 0,\pm \pi.$ 

\begin{figure}[H] 
\centering
\subfigure[ $\overline{P}_{\infty}(\psi_R),\overline{P}_{\infty}(\psi_U)$]{\includegraphics[height=6 cm,width=6.5 cm]{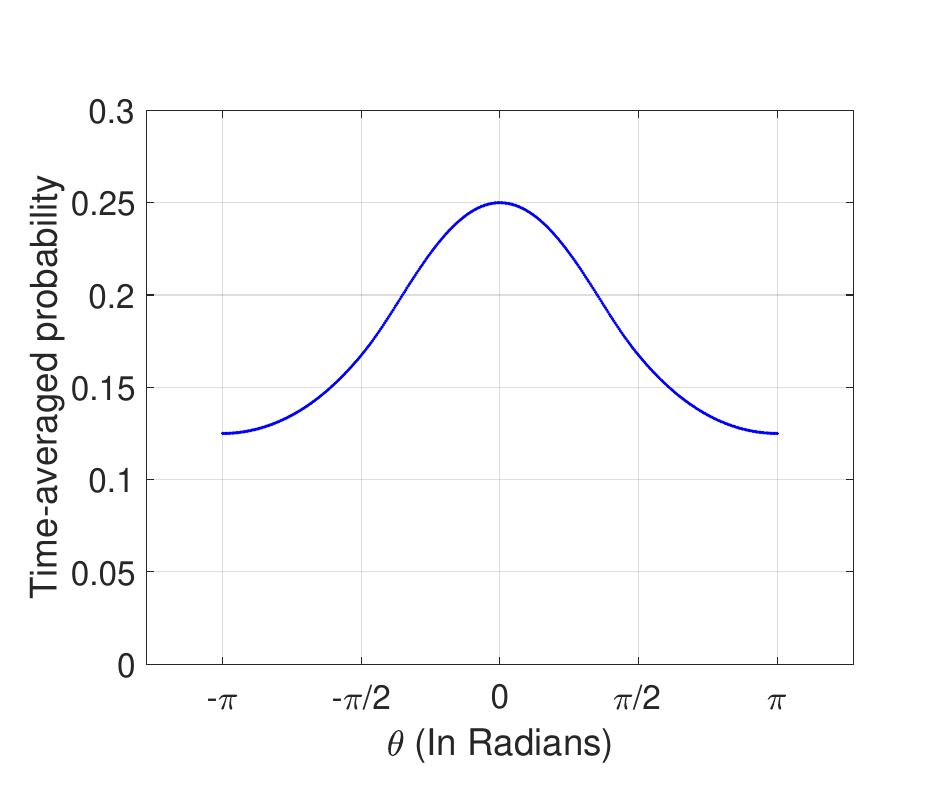}} 
\hspace{0.3cm}
\subfigure[ $\overline{P}_{\infty}(\psi_L),\overline{P}_{\infty}(\psi_D)$]{\includegraphics[height=5.7 cm,width=6.5 cm]{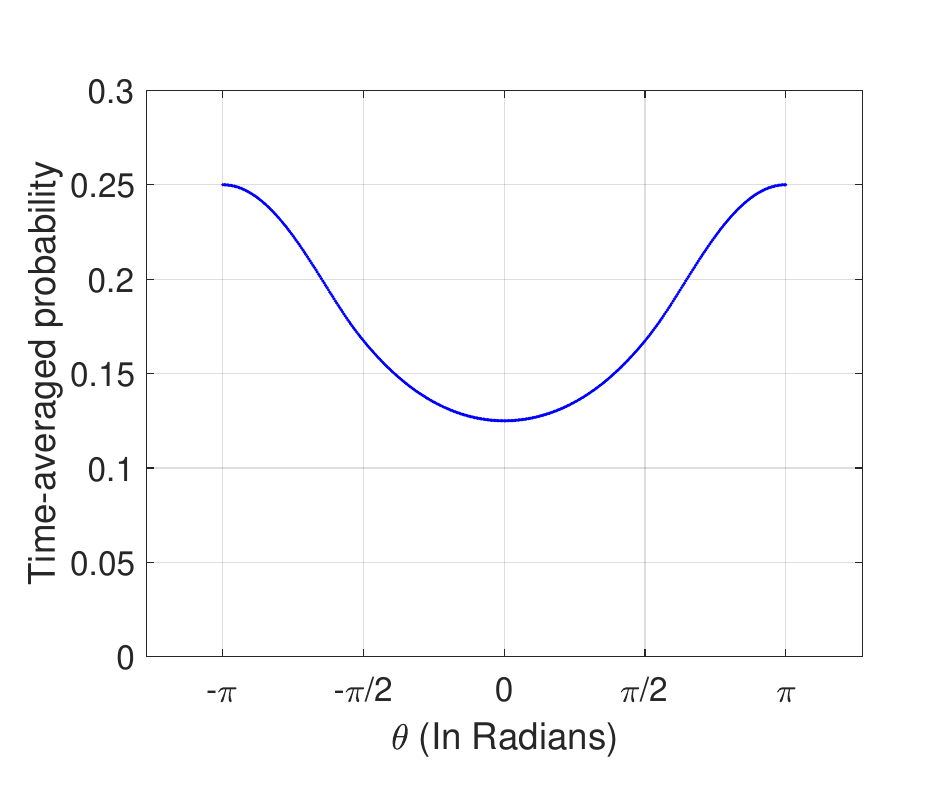}}
\caption{Numerical values of $\overline{P}_\infty ((0,0);\psi_S(0)), S\in\{R, L, U, D\}$ when the coins belong to $Y_\theta$ with $400$  equidistant values of $\theta$ in $[-\pi, \pi].$  } \label{fig1}
\end{figure}

Following a similar process, for $C\in Y_\theta,\theta =0,\pm \pi$ we evaluate that $\overline{P}_\infty ((0,0);\psi_S(0))=\frac{1}{8}$ for $\theta= \pi,-\pi,S\in \{R,U\}$ or $\theta=0,S\in \{L,D\},$ and $\overline{P}_\infty ((0,0);\psi_S(0))=\frac{1}{4}$ for $\theta=0,S\in \{R,U\}$ or $\theta= \pi,-\pi,S\in \{L,D\}.$ In Figure \ref{fig1} we plot $\overline{P}_\infty ((0,0);\psi_S(0))$ for different values of $\theta$ obtained by discretizing the interval $[-\pi, \pi]$ into $400$ equidistant points. 
Note that, if 
$\ket{\psi_S(0)}=[\psi_j(0)]$ then $\psi_j(0)=1$ if $j=l(S)+2N^2-2$ and $0$ otherwise. 

It follows from Figure \ref{fig1} that the values of $\overline{P}_\infty ((0,0);\psi_S(0)),S\in\{R,L,U,D\}$ corresponding to different $\theta$ are symmetric with respect to the vertical axis passing through $\theta=0.$ Finally, we conclude that the proposed DTQWs with canonical initial coin states on infinite lattice localize when the initial position is $(0,0).$ A similar analysis can be performed for any vertex $(x,y)\in \mathbb{Z} \times \mathbb{Z}$ as an initial position.

\begin{remark}{(Initial state correspond to $\overline{P}_{\infty}(S',(0,0);\psi(0))=0$)} \label{remark:infi N}
For the proposed quantum walks with an initial state $\ket{\psi(0)},$ using (\ref{coeffi:alpha_S}), (\ref{eqn:formulac}) and (\ref{def:finalp}) we obtain \beano\overline{P}_{\infty}(S',(0,0);\psi(0))&=& \lim_{N \rightarrow \infty}\frac{1}{N^4} \left(\left|\sum_{j=1}^{4N^2} C_{r,j,1}\right|^2+\left|\sum_{j=1}^{4N^2} C_{r,j,2}\right|^2\right) \nonumber \\
&=&\lim_{N \rightarrow \infty}\frac{1}{N^4}\left(\left|\sum_{n=1}^{\frac{N-3}{2}}\sum_{m=n+1}^{\frac{N-1}{2}} \sum_{j=1}^{4N^2}c_{r,j,n,m,1}\right|^2+\left|\sum_{n=1}^{\frac{N-3}{2}}\sum_{m=n+1}^{\frac{N-1}{2}}\sum_{j=1}^{4N^2} c_{r,j,n,m,2}\right|^2\right)\nonumber\\
&=&\lim_{N \rightarrow \infty}\left(\left|\sum_{n=0}^{N-1}\sum_{m=0}^{N-1} \eta_{r,n,m,1}\bra{\eta_{n,m,1}}\ket{\psi(0)}\right|^2+\left|\sum_{n=0}^{N-1}\sum_{m=0}^{N-1} \eta_{r,n,m,2}\bra{\eta_{n,m,2}}\ket{\psi(0)}\right|^2\right),
\eeano
$r=l(S')+2N^2-2.$
Then it is clear that if the initial state $\ket{\psi(0)}$ is orthogonal to the eigenvectors $\ket{\eta_{n,m,1}},\ket{\eta_{n,m,2}}$  corresponding to the eigenvalues $-1$ and $1$ of $\U,$ $0\leq n,m\leq N-1$, or in other words, if the eigenvectors $\ket{v_{n,m,1}},\ket{v_{n,m,2}}$ are orthogonal to $\ket{\psi_{0,0}(0)}$ for all $(n,m),$ then $\overline{P}_{\infty}(S',(0,0);\psi(0))=0$ for $S'\in \{R,L,U,D\}$ and consequently $\overline{P}_{\infty}((0,0);\psi(0))=0.$
Thus for such initial states, the walk does not localize at $(0,0).$  A similar analysis can be done for any vertex $(x,y).$
\end{remark}

 Now, in the following propositions we determine initial coin states which provide zero value of the total time-averaged probability.

\begin{proposition}\label{pro:ini st Y}
 The walk with $C\in Y_\theta,\theta\in [-\pi,\pi],$ does not show localization at the initial position $(0,0)\in \mathbb{Z}\times \mathbb{Z}$ if and only if the initial coin state $\ket{\psi_{0,0}(0)}\in I^{(y)}_0$ where
 $$I^{(y)}_0=\left\{[a,b,c,d]^T\in \mathbb{C}^4 \, |\, a=-c=\alpha(1+\cos \theta),b=-d=-\alpha\sin \theta,\alpha\in \mathbb{C},|\alpha|^2=\frac{1}{4 (1+\cos \theta)}\right\}$$ if $\theta\notin\{0,\pm \pi\},$ $I^{(y)}_0=\{[\alpha,0,-\alpha,0]^T|\alpha\in \mathbb{C},|\alpha|^2=1/2\}$ if $\theta=0,$ and  $I^{(y)}_0=\{[0,\alpha,0,-\alpha]^T | \alpha\in \mathbb{C},|\alpha|^2=1/2\},$ if $\theta=\pm \pi.$
\end{proposition}

\pf The `if' part is easy to verify. 
 The `only if' part follows from Remark \ref{remark:infi N} that the initial coin state $\ket{\psi_{0,0}(0)}$ for which the underlying walk does not exhibit localization at $(0,0)$ can be determined by considering that the following two conditions are satisfied $$\langle v_{n,m,1}\vert \psi_{0,0}(0)\rangle=\langle v_{n,m,2}\vert \psi_{0,0}(0)\rangle=0,$$ for each $(n,m),$ $0\leq n,m\leq N-1$. Thus by evaluating $\ket{v_{n,m,1}},\ket{v_{n,m,2}}$ from Lemma \ref{lemma:eig u_(n,m)} and Lemma \ref{lemma:eig u_(n,m) per}, the desired results follows after performing some algebraic calculations. $\hfill{\square}$
 
 Clearly, by choosing initial state  $\ket{\psi_S(0)},S\in\{R, L, U, D\}$ for the walk with $C\in Y_\theta,\theta \in [-\pi,\pi]$ we obtain by Proposition \ref{pro:ini st Y} that $\overline{P}_\infty ((0,0);\psi_S(0))\neq 0,$ which can be observed in Figure \ref{fig1}.



\begin{figure}[H] 
\centering
\subfigure[$\ket{\psi_{0,0}(0)}\not \in I^{(y)}_0,N=31.$]{\includegraphics[height=4 cm,width=5 cm]{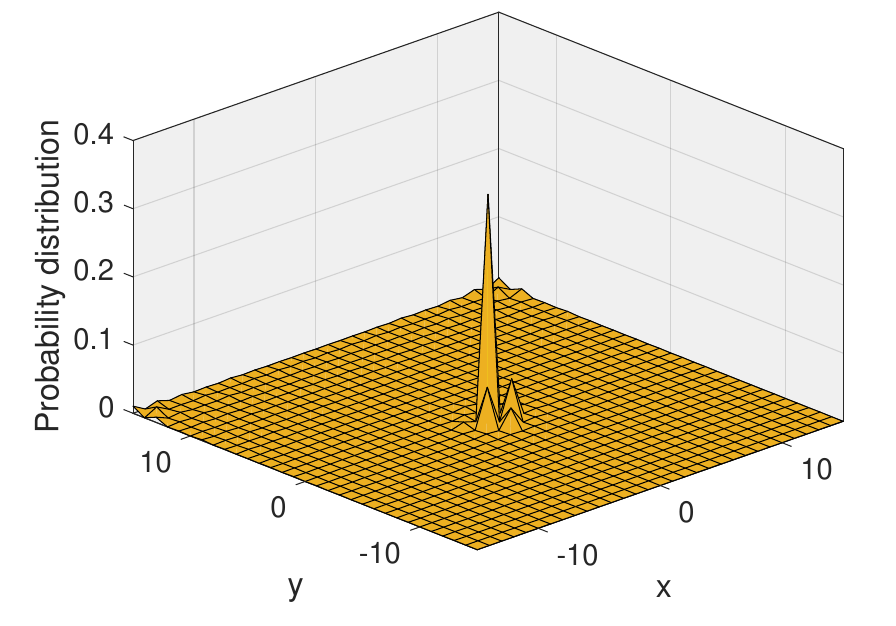}} 
\subfigure[$\ket{\psi_{0,0}(0)} \in I^{(y)}_0, N=41.$ ]{\includegraphics[height=4.2 cm,width=5.1 cm]{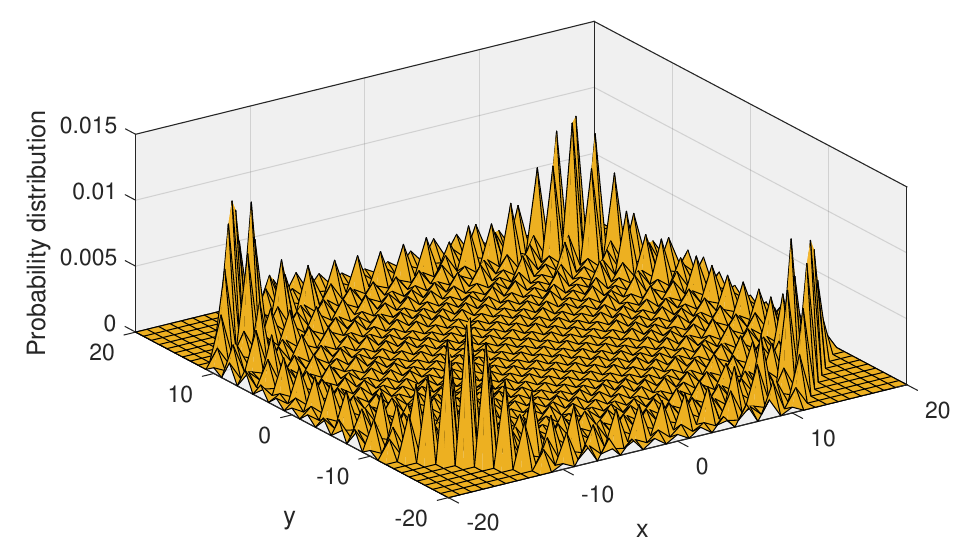}}
\caption{$(a)$ and $(b)$ show the probability distribution of the Grover walk after $t=30$ time steps,  with initial state vectors $[1,0,0,0]^T$ and $[-1/2,1/2,1/2,-1/2]^T\in I^{(y)}_0,$ respectively.  Here we choose the system size $N=31$ and $N=41$ for $(a)$ and $(b)$ respectively.
}\label{fig:2d Grover}
\end{figure}

\begin{figure}[H] 
\centering
\subfigure[ $\ket{\psi_{0,0}(0)}\not\in I^{(y)}_0,N=31$]{\includegraphics[height=4.3 cm,width=5 cm]{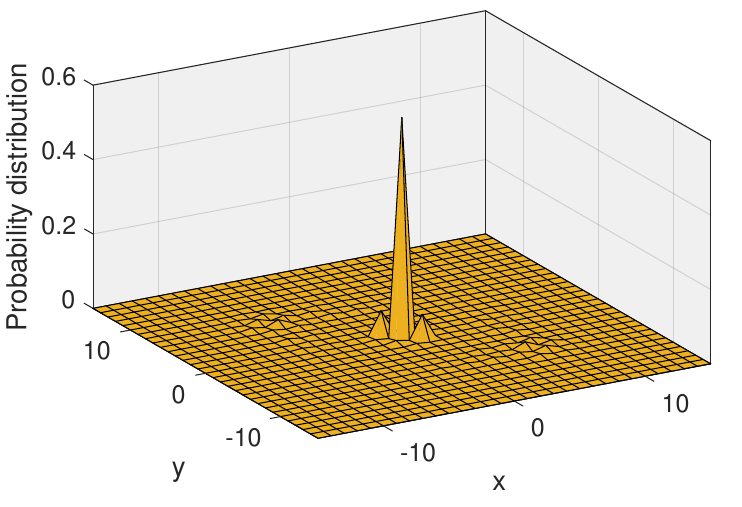}} 
\hspace{0.4 cm}
\subfigure[$\ket{\psi_{0,0}(0)} \in I^{(y)}_0,N=31$]{\includegraphics[height=4.5 cm,width=5.4 cm]{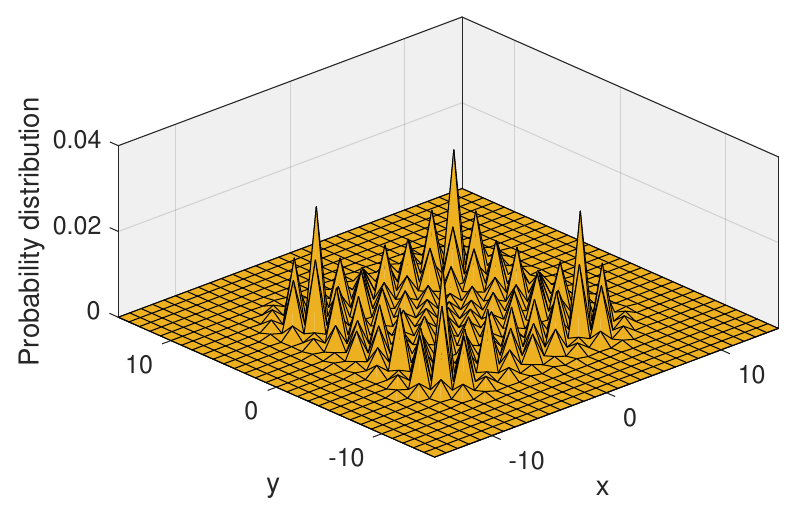}}
\caption{$(a)$ and $(b)$ show the probability distribution of the walk with $C\in Y_\theta$ for $\theta=\pi/6,$ after $t=30$ time steps, with initial coin state vectors $[1/2,1/2,1/2,1/2]^T$ and $\left[\frac{1}{2 \sqrt{4+2 \sqrt 3}}(2+\sqrt 3),-1,-(2+\sqrt 3),1\right]^T\in I^{(y)}_0,$ respectively.  Here we choose $N=31$ for both the cases $(a)$ and $(b).$}\label{fig:2d other}
\end{figure}

 The theoretical results of Proposition \ref{pro:ini st Y} regarding the localization at $(0,0)$ are illustrated numerically for some finite system size
 $N$ in Figure \ref{fig:2d Grover} and Figure \ref{fig:2d other} for the Grover walk and the walk with coin $C\in Y_\theta, \theta=\pi/6,$ respectively. In each case we plot the probability distribution $ P_t((x,y);\psi(0))$ after $t=30$ time steps for two different types of initial coin states, one of which does not belong to $I_0^{(y)}$ and another is taken from $I_0^{(y)}.$ When $\ket{\psi_{0,0}(0)}\notin I_0^{(y)},$ we see that there are peaks in the probability distribution at the origin $(0,0),$ while for $\ket{\psi_{0,0}(0)}\in I_0^{(y)},$ the central peaks vanish and the probabilities spread symmetrically towards the other vertices. 



\subsection{With coins from $X_\theta$ and $Z_\theta$}

In this section, we consider the proposed walks when the coin operator belongs to $X_{\theta}$ or $Z_\theta,$ $[-\pi, \, \pi].$
If the coin operator $C\in X_\theta$ then the walk evolution operator  $${\U'}_{n,m}=D_{n,m}C=
\bmatrix{\frac{1}{2}\sin\theta\omega^{-n}&-\frac{1}{2}\sin\theta\omega^{-n}&\frac{1}{2}(1+\cos\theta)\omega^{-n}&\frac{1}{2}(1-\cos\theta)\omega^{-n}\\-\frac{1}{2}\sin\theta\omega^{n}&\frac{1}{2}\sin\theta\omega^{n}&\frac{1}{2}(1-\cos\theta)\omega^{n}&\frac{1}{2}(1+\cos\theta)\omega^{n}\\\frac{1}{2}(1-\cos\theta)\omega^{-m}&\frac{1}{2}(1+\cos\theta)\omega^{-m}&\frac{1}{2}\sin\theta\omega^{-m}&-\frac{1}{2}\sin\theta\omega^{-m}\\\frac{1}{2}(1+\cos\theta)\omega^{m}&\frac{1}{2}(1-\cos\theta)\omega^{m}&-\frac{1}{2}\sin\theta\omega^{m}&\frac{1}{2}\sin\theta\omega^{m}},$$
and if the coin operator $C\in Z_\theta$ then the walk evolution operator

 $${\U''}_{n,m}=D_{n,m}C=
\bmatrix{\frac{1}{2}\sin\theta\omega^{-n}&\frac{1}{2}(1+\cos\theta)\omega^{-n}&\frac{1}{2}(1-\cos\theta)\omega^{-n}&-\frac{1}{2}\sin\theta\omega^{-n}\\\frac{1}{2}(1-\cos\theta)\omega^{n}&\frac{1}{2}\sin\theta\omega^{n}&-\frac{1}{2}\sin\theta\omega^{n}&\frac{1}{2}(1+\cos\theta)\omega^{n}\\\frac{1}{2}(1+\cos\theta)\omega^{-m}&-\frac{1}{2}\sin\theta\omega^{-m}&\frac{1}{2}\sin\theta\omega^{-m}&\frac{1}{2}(1-\cos\theta)\omega^{-m}\\-\frac{1}{2}\sin\theta\omega^{m}&\frac{1}{2}(1-\cos\theta)\omega^{m}&\frac{1}{2}(1+\cos\theta)\omega^{m}&\frac{1}{2}\sin\theta\omega^{m}},$$ 
where $\theta \in [-\pi,\pi]$ and $D_{n,m}=diag(\omega^{-n},\omega^{n},\omega^{-m},\omega^{m}),$  $\omega=e^{2\pi i/N},\,\,m,n\in\{0,\ldots,N-1\}.$


First we determine the eigenpairs of the unitary operators $\U'_{n,m}$ and  $\U''_{n,m}$ as follows.

\begin{lemma} \label{lem:eig X}
A set of eigenpairs $(\lambda_{n,m,k},\ket{v'_{n,m,k}}),\,k=1,2,3,4$ of ${\U'}_{n,m}$ are as follows. For $\theta \neq 0, \pm \pi,$
\beano && \lambda_{n,m,1}=-1, \lambda_{n,m,2}= 1, 
 \lambda_{n,m,3}=\frac{\sin{\theta}(\cos{\zeta_n}+\cos{\zeta_m})-i\sqrt{4-\sin^2{\theta}(\cos{\zeta_n}+\cos{\zeta_m})^2}}{2}, \\
&& \lambda_{n,m,4} = \frac{\sin{\theta}(\cos{\zeta_n}+\cos{\zeta_m})+i\sqrt{4-\sin^2{\theta}(\cos{\zeta_n}+\cos{\zeta_m})^2}}{2},
\eeano
\begin{eqnarray*} \ket{v'_{n,m,k}}= && \left[ \left(\dfrac{\sin{\theta}-(1-\cos{\theta})\lambda_{n,m,k}\omega^{-n}}{(1-\cos{\theta})-\sin{\theta}\lambda_{n,m,k}\omega^m}\right)\left(\dfrac{1+\cos{\theta}-\lambda_{n,m,k}\sin{\theta}\omega^m}{\sin{\theta}-(1+\cos{\theta})\lambda_{n,m,k}\omega^{n}}\right), \right.
 1, \,\,\\
&& \hfill{\left.-\dfrac{\sin{\theta}-(1-\cos{\theta})\lam_{n,m,k}\omega^{-n}}{(1-\cos{\theta})-\lambda_{n,m,k}\sin{\theta}\omega^{m}}, \,\, 
-\dfrac{\sin{\theta}-(1+\cos{\theta})\lambda_{n,m,k}\omega^{-n}}{(1+\cos{\theta})-\lam_{n,m,k}\sin{\theta}\omega^{-m}}\right]^T};\end{eqnarray*}
for $\theta=0,$
\beano 
& \lambda_{n,m,1}=-1, \lambda_{n,m,2}= 1, \lambda_{n,m,3}=-i,\lambda_{n,m,4} =i,\\
&\ket{v'_{n,m,k}}=\bmatrix{\lambda_{n,m,k}\omega^{-m}, \,\,\lambda^3_{n,m,k}\omega^{n}, \,\, \lambda^2_{n,m,k}\omega^{n-m},\,\, 1}^T;
\eeano
and for $\theta=\pm \pi,$
\beano &\lambda_{n,m,1}=-1, \lambda_{n,m,2}= 1, \lambda_{n,m,3}=-i,\lambda_{n,m,4} =i ,\\
&\ket{v'_{n,m,k}}=\bmatrix{\lambda^3_{n,m,k}\omega^{-n}, \,\,\lambda_{n,m,k}\omega^{-m}, \,\, \lambda^2_{n,m,k}\omega^{-n-m},\,\, 1}^T;
\eeano
where $\omega=e^{2\pi i/N}.$
\end{lemma}
\pf   
The proof follows from the fact that the characteristic polynomial of ${\U'}_{n,m}$ is $$\chi_{\mathrm{U'}_{n,m}}(\lambda)=\lambda^4-\sin{\theta}(\cos{\zeta_m}+\cos{\zeta_n})\lambda^3+\sin{\theta}(\cos{\zeta_m}+\cos{\zeta_n})\lambda-1.$$ $\hfill{\square}$

\begin{lemma}\label{lem:eig Z}
A set of eigenpairs $(\lambda_{n,m,k},\ket{v''_{n,m,k}}),k=1,2,3,4$ of ${\U''}_{n,m}$ are given as follows. For $\theta \neq 0, \pm \pi,$
\beano && \lambda_{n,m,1}=-1, \lambda_{n,m,2}= 1, 
 \lambda_{n,m,3}=\frac{\sin{\theta}(\cos{\zeta_n}+\cos{\zeta_m})-i\sqrt{4-\sin^2{\theta}(\cos{\zeta_n}+\cos{\zeta_m})^2}}{2}, \\
&& \lambda_{n,m,4} = \frac{\sin{\theta}(\cos{\zeta_n}+\cos{\zeta_m})+i\sqrt{4-\sin^2{\theta}(\cos{\zeta_n}+\cos{\zeta_m})^2}}{2}, 
\eeano
\beano \ket{v''_{n,m,k}}= &&\left[ \left(\dfrac{\sin{\theta}-(1+\cos{\theta})\lambda_{n,m,k}\omega^m}{(1+\cos{\theta})-\lambda_{n,m,k}sin{\theta}\omega^{n}}\right),
 -\left(\dfrac{(1-\cos{\theta})-\lambda_{n,m,k}\sin{\theta}\omega^{n}}{\sin{\theta}-(1-\cos{\theta})\lambda_{n,m,k}\omega^{-n}}\right)\left(\dfrac{\sin{\theta}-(1+\cos{\theta})\lambda_{n,m,k}\omega^m}{(1+\cos{\theta})-\lambda_{n,m,k}sin{\theta}\omega^{n}}\right),\right. \\
&& \hfill{\left.1,
\left(-\dfrac{(1+\cos{\theta})-\sin{\theta}\lam_{n,m,k}\omega^{m}}{\sin{\theta}-(1+\cos{\theta})\lam_{n,m,k}\omega^{-m}}\right)\right]^T};\eeano
for $\theta=0,$
\beano 
& \lambda_{n,m,1}=-1, \lambda_{n,m,2}= 1, \lambda_{n,m,3}=-i,\lambda_{n,m,4} =i,\\
&\ket{v''_{n,m,k}}=\bmatrix{\lambda^2_{n,m,k}, \,\,\lambda^3_{n,m,k}\omega^{n}, \,\, \lambda_{n,m,k}\omega^{-m},\,\, 1}^T;
\eeano
and for $\theta=\pm \pi,$
\beano &\lambda_{n,m,1}=-1, \lambda_{n,m,2}= 1, \lambda_{n,m,3}=-i,\lambda_{n,m,4} =i ,\\
&\ket{v''_{n,m,k}}=\bmatrix{\lambda^2_{n,m,k}\omega^{-n-m}, \,\,\lambda_{n,m,k}\omega^{-m}, \,\, \lambda^3_{n,m,k}\omega^{-m},\,\, 1}^T;
\eeano
where $\omega=e^{2\pi i/N}.$ 
\end{lemma}
\pf   
The proof follows from the fact that the characteristic polynomial of ${\U''}_{n,m}$ is $$\chi_{\mathrm{U''}_{n,m}}(\lambda)=\lambda^4-\sin{\theta}(\cos{\zeta_m}+\cos{\zeta_n})\lambda^3+\sin{\theta}(\cos{\zeta_m}+\cos{\zeta_n})\lambda-1.$$ $\hfill{\square}$

Note that $\Lambda(\U_{n,m})=\Lambda(\U'_{n,m})=\Lambda(\U''_{n,m}).$ Besides, a set of orthonormal eigenvectors of the evolution matrix corresponding to the DTQWs defined by generalized Grover coins in $X_\theta$ and $Z_\theta$ can be obtained by employing similar arguments as in the previous section. Thus an integration formula of time-average probabilities can be determined for these walks. Indeed,  recall from (\ref{def:finalp}) that
\beano
\overline{P}_\infty(S',(0,0);\psi_S(0))
&=&\lim_{N \rightarrow \infty}\frac{1}{N^4}\left(\left|\sum_{n=1}^{(N-3)/2}\sum_{m=n+1}^{(N-1)/2} c_{r,j,n,m,1}\right|^2+\left|\sum_{n=1}^{(N-3)/2}\sum_{m=n+1}^{(N-1)/2} c_{r,j,n,m,2}\right|^2\right),
\eeano
$S'\in \{R,L,U,D\}.$
The values of $c_{r,j,n,m,k},k=1,2$ can be calculated using equation (\ref{eqn:formulac}).

It may be noted that explicit expressions of $c_{r,j,n,m,k},k=1,2$ for different $r$ and $j$ is complicated. Thus we perform numerical computations for the values of $\overline{P}_\infty ((0,0);\psi_S(0))$ for different values of $\theta.$  We find that $c_{r,j,n,m,k}$ are same for some pairs $(r,j),$ for example $c_{1,3,n,m,k}=c_{1,4,n,m,k}=c_{2,3,n,m,k}=c_{2,4,n,m,k},\,c_{1,2,n,m,k}=c_{3,4,n,m,k},\,c_{1,1,n,m,k}=c_{2,2,n,m,k}=c_{3,3,n,m,k}=c_{4,4,n,m,k}$ which finally results in   $\overline{P}_\infty ((0,0);\psi_S(0))$ to be same for all $S\in \{R,L,U,D\}$ since $c_{r,j,n,m,k}=c_{j,r,n,m,k},$ when the coin operator belongs to $X_\theta.$ 

\begin{figure}[H]
    \centering
    \includegraphics[height=5.5 cm,width=5.5 cm]{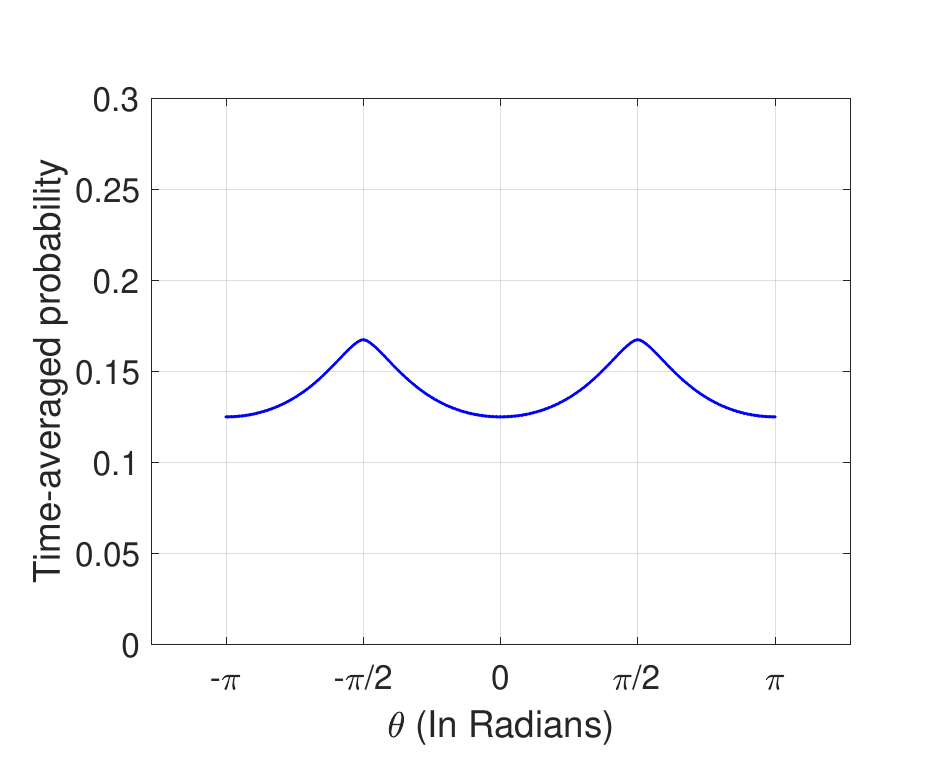}
   \caption{ Numerical values of $\overline{P}_\infty ((0,0);\psi_S(0)), S\in\{R, L, U, D\}$ when the coins belong to $X_\theta$ with $400$  equidistant values of $\theta$ in $[-\pi, \pi].$ } \label{fig2}
\end{figure}

We plot $\overline{P}_\infty ((0,0);\psi_R(0))$ considering $400$ equidistant values of $\theta$ in the interval $[-\pi, \pi]$ in Figure \ref{fig2}. It can be further observed from Figure \ref{fig2} that the probability distribution $\overline{P}_\infty ((0,0);\psi_R(0))$ is symmetric with respect to the vertical axis through $\theta=0.$ Probabilities are maximum at $\theta=\pi/2, -\pi/2$ i.e. coin operators are $G,P_{(12)(34)}G$ and minimum at $\theta=0, \pi,-\pi$ i.e. coin operators are $P_{(1324)},P_{(1423)}$. Finally, we conclude the proposed walks corresponding to these values of the coin parameter $\theta$ with the initial state $\psi_S(0),S\in \{R,L,U,D\}$ localize at the initial position $(0,0)$ of the infinite lattice.

In the next proposition we show that the proposed walks with $C\in X_\theta, \theta\in [-\pi,\pi], \theta\neq -\frac{\pi}{2}$ i.e. $C\neq G,$ show localization at initial position for any initial coin state. 
\begin{proposition}
 The walk with $C\in X_\theta,\theta\in [-\pi,\pi], \theta\neq -\frac{\pi}{2}$ shows localization at the position $(0,0)\in \mathbb{Z}\times \mathbb{Z}$ for any initial coin state. 
\end{proposition}
\pf First consider the case for $\theta=0.$ Then by Lemma \ref{lem:eig X} the eigenvectors of $\U'_{n,m}$ corresponding to the eigenvalues $-1$ and $1$ are \beano \ket{v'_{n,m,1}} = \bmatrix{-\omega^{-m}, \,\,-\omega^{n}, \,\, \omega^{n-m},\,\, 1}^T \,\,\mbox{and}\,\,
\ket{v'_{n,m,2}} =\bmatrix{\omega^{-m}, \,\,\omega^{n}, \,\, \omega^{n-m},\,\,1}^T.\eeano
Next for $\theta=\pm \pi$ we have
 \beano \ket{v'_{n,m,1}} =\bmatrix{-\omega^{-n}, \,\,-\omega^{-m}, \,\, \omega^{-n-m},\,\, 1}^T \,\,\mbox{and}\,\,
\ket{v'_{n,m,2}} =\bmatrix{\omega^{-n}, \,\,\omega^{-m}, \,\, \omega^{-n-m},\,\, 1}^T.\eeano
Finally, for $\theta\neq 0,\pm \pi, -\frac{\pi}{2}$ we recall  $\ket{v'_{n,m,1}}$ and $\ket{v'_{n,m,2}}$ from Lemma \ref{lem:eig X}.
Then by using Remark \ref{remark:infi N}, the proof follows from the fact that there does not exist any initial coin state which is orthogonal to both $\ket{v'_{n,m,1}}$ and $\ket{v'_{n,m,2}}$ for each $n,m$ where $0\leq n,m \leq N-1.$
$\hfill{\square}$

\begin{figure}[H] 
\centering
\subfigure[ $\overline{P}_{\infty}(\psi_R),\overline{P}_{\infty}(\psi_D)$]{\includegraphics[height=6 cm,width=6.5 cm]{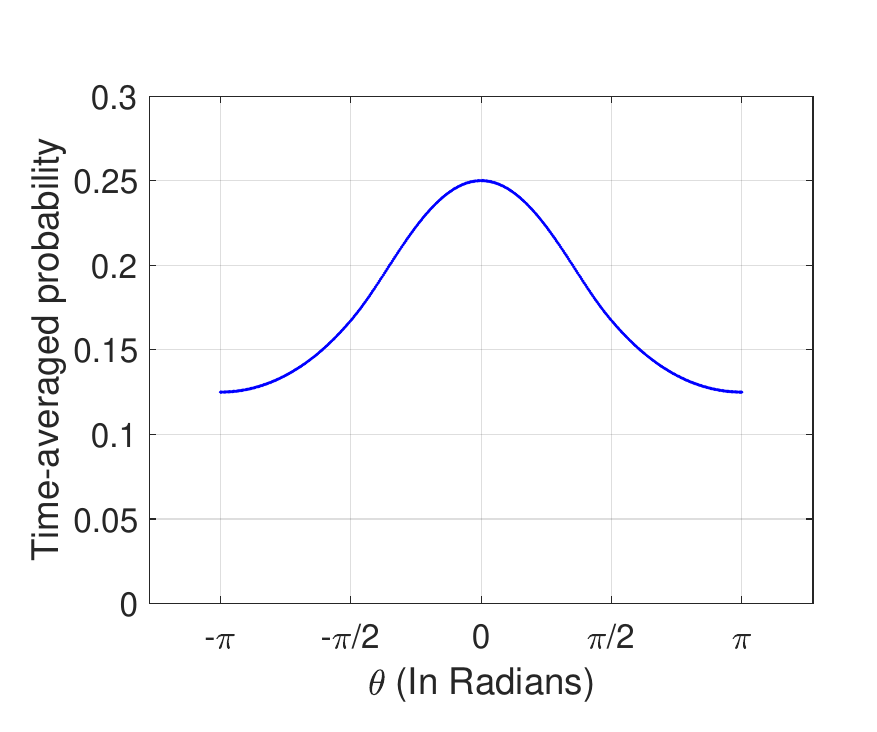}} 
\hspace{0.3cm}
\subfigure[ $\overline{P}_{\infty}(\psi_L),\overline{P}_{\infty}(\psi_U)$]{\includegraphics[height=6 cm,width=6.5 cm]{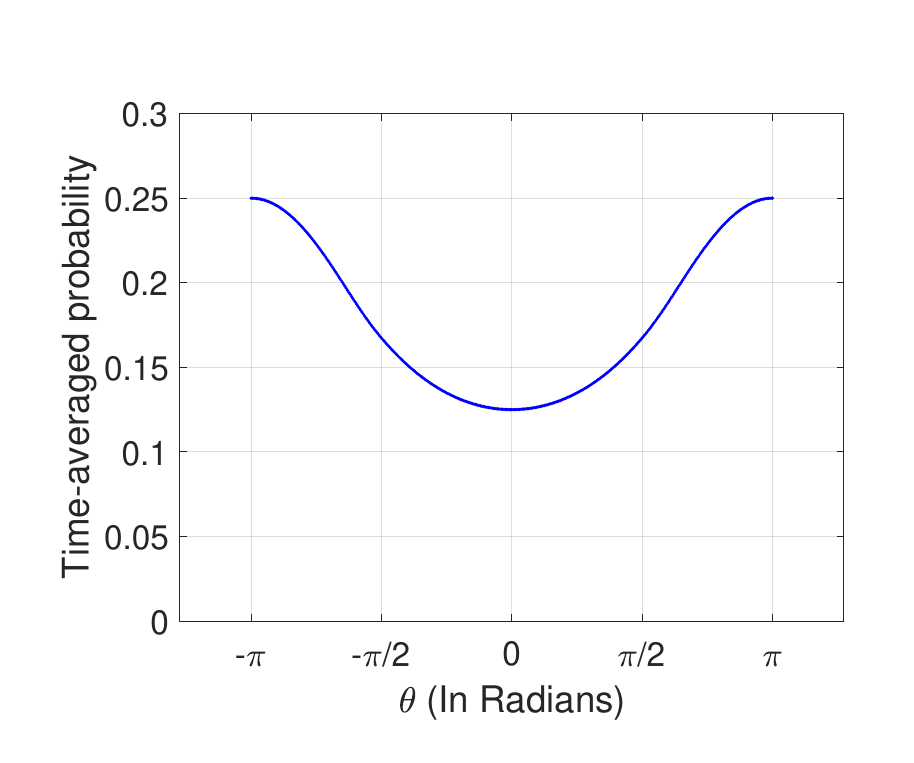}}
\caption{Numerical values of $\overline{P}_\infty ((0,0);\psi_S(0)), S\in\{R, L, U, D\}$ when the coins belong to $Z_\theta$ for $400$  equidistant values of $\theta$ in $[-\pi, \pi].$ 
}\label{fig3}
\end{figure}

Next we plot the total time-averaged probabilities in Figure \ref{fig3} for canonical initial coin states when the coin operator is in $Z_\theta,$ for different values of $\theta$ obtained by discretizing the interval $[-\pi, \pi]$ into $400$ equidistant points. Observe that,
$\overline{P}_\infty ((0,0);\psi_S(0)), S\in\{R,L,U,D\}$ are symmetric with respect to the vertical axis $\theta=0$ and the time-averaged probability decreases or increases depending on the initial coin state when $|\theta|$ increases.  Obviously, the DTQWs defined by these generalized Grover coins with the canonical initial coin states localize at the vertex $(0,0)$ of the infinite lattice.

In following proposition we determine all the initial coin states for which the walks do not exhibit localization when the coin operator $C\in Z_\theta.$


\begin{proposition}\label{pro:ini st Z per}
 The walk with $C\in Z_\theta,\theta\in [-\pi,\pi],$ does not show localization at the initial position $(0,0)\in \mathbb{Z}\times \mathbb{Z}$ if and only if the initial coin state $\ket{\psi_{0,0}(0)}\in I^{(z)}_0$ where
 $$I_0^{(z)}=\left\{[a,b,c,d]^T\in \mathbb{C}^4|a=-d=\alpha(1+\cos \theta),b=c=-\alpha\sin \theta,\alpha\in \mathbb{C},|\alpha|^2=\frac{1}{4 (1+\cos \theta)}\right\}$$ if $\theta\notin\{0,\pm \pi\},$ $I^{(z)}_0=\{[\alpha,0,0,-\alpha]^T|\alpha\in \mathbb{C},|\alpha|^2=1/2\}$ if $\theta=0,$ and  $I^{(z)}_0=\{[0,\alpha,-\alpha,0]^T | \alpha\in \mathbb{C},|\alpha|^2=1/2\},$ if $\theta=\pm \pi.$
\end{proposition}
\pf The proof follows similarly as Propositions \ref{pro:ini st Y} by using Lemma \ref{lem:eig Z}. $\hfill{\square}$


Note that, by Proposition \ref{pro:ini st Z per}, if the walk is defined for the coin $C\in Z_\theta$ and the initial state  is $\ket{\psi_S(0)},S\in\{R, L,U,D\},$ we obtain $\overline{P}_\infty ((0,0);\psi_S(0))\neq 0,$ $\theta \in [-\pi,\pi].$

\subsection{With coins from $W_\theta$}

In this section, we consider the proposed walks when the coin operator lies in  $W_{\theta},\theta \in [-\pi, \, \pi].$ As mentioned before, note that $W_\theta$ does not contain the Grover matrix and hence this set of parametric coins need not be treated as continuous deformation of the Grover matrix. We demonstrate that the total probability value of these walks with initial coin state $\ket{\psi_S(0)}$ approaches to zero when  $S\in\{R,L\}$ and $\theta\rightarrow 0,$ and  $S\in\{U,D\}$ and $\theta\rightarrow \pm\pi.$ 




Let $C\in W_\theta.$ Then the walk evolution operator is given by
$$\U'''_{n,m}=D_{n,m}C=\bmatrix{\frac{1}{2}(1+\cos \theta)\omega^{-n} & \frac{1}{2}(1-\cos \theta)\omega^{-n} & \frac{1}{2}\sin \theta\omega^{-n} & -\frac{1}{2}\sin \theta\omega^{-n}\\ \frac{1}{2}(1-\cos \theta)\omega^{n} & \frac{1}{2}(1+\cos \theta)\omega^{n} & -\frac{1}{2}\sin \theta\omega^{n} & \frac{1}{2}\sin \theta\omega^{n}\\ \frac{1}{2}\sin \theta\omega^{-m} & -\frac{1}{2}\sin \theta\omega^{-m} & \frac{1}{2}(1-\cos \theta)\omega^{-m} & \frac{1}{2}(1+\cos \theta)\omega^{-m}\\-\frac{1}{2}\sin \theta\omega^{m} & \frac{1}{2}\sin \theta\omega^{m} & \frac{1}{2}(1+\cos \theta)\omega^{m} & \frac{1}{2}(1-\cos \theta)\omega^{m}}$$
for $-\pi \leq \theta \leq  \pi, \omega=e^{2\pi i/N},\,\,m,n\in\{0,\ldots,N-1\}.$
Now we determine a set of eigenpairs of $\U'''_{n,m}$ in the following lemma.

\begin{lemma} \label{lem:eig W}
A set of eigenpairs $(\lambda_{n,m,k},\ket{v'''_{n,m,k}}),k=1,2,3,4$ of ${\U''}_{n,m}$ are given as follows.
 For $\theta \neq 0,\pm \pi$,
$$\lambda_{n,m,1}=-1,\,\lambda_{n,m,2}=1,\,
\lambda_{n,m,3}=e^{i{\eta}},\,\lambda_{n,m,4}=e^{-i{\eta}},$$ where  $\cos{\eta}=\frac{(1+\cos{\theta})\cos{\zeta_n}+(1-\cos{\theta})\cos{\zeta_m}-1}{2}$,
 \beano\ket{v'''_{n,m,k}}&= &\left[ -\dfrac{\sin{\theta}(1-\lam_{n,m,k}\omega^{m})}{(1+\cos{\theta})(1-\lam_{n,m,k}\omega^{-n})},
 -\dfrac{\sin{\theta}(1-\lam_{n,m,k}\omega^{m})}{(1+\cos{\theta})(1-\lam_{n,m,k}\omega^{-n})},1,
-\dfrac{1-\lam_{n,m,k}\omega^m}{1-\lam_{n,m,k}\omega^{-m}}\right]^T;\eeano
for $\theta=0,$ \beano 
& \lambda_{n,m,1}=-1, \lambda_{n,m,2}= 1, \lambda_{n,m,3}=\omega^{-n},\lambda_{n,m,4}=\omega^{n},\\
&\ket{v'''_{n,m,1}}=\bmatrix{0, \,\,0, \,\, -\omega^{-m},\,\,1}^T,\,\,\ket{v'''_{n,m,2}}=\bmatrix{0, \,\,0, \,\, \omega^{-m},\,\, 1}^T,\\
&\ket{v'''_{n,m,3}}=\bmatrix{1, \,\,0, \,\, 0,\,\, 0}^T,\,\,\ket{v'''_{n,m,4}}=\bmatrix{0, \,\,1, \,\, 0,\,\, 0}^T;
\eeano
and for $\theta=\pm \pi,$ \beano 
& \lambda_{n,m,1}=-1, \lambda_{n,m,2}= 1, \lambda_{n,m,3}=\omega^{-m},\lambda_{n,m,4} =\omega^{m},\\
&\ket{v'''_{n,m,1}}=\bmatrix{-\omega^{-n}, \,\,1, \,\, 0,\,\, 0}^T,\,\,\ket{v'''_{n,m,2}}=\bmatrix{\omega^{-n}, \,\,1, \,\, 0,\,\, 0}^T,\\
&\ket{v'''_{n,m,3}}=\bmatrix{0, \,\,0, \,\, 1,\,\, 0}^T,\,\,\ket{v'''_{n,m,4}}=\bmatrix{0, \,\,0, \,\, 0,\,\, 1}^T;
\eeano
where $\omega=e^{2\pi i/N}.$ 
\end{lemma}
\pf The characteristic polynomial of ${\U'''}_{n,m}$ is $$\chi_{\mathrm{U'''}_{n,m}}(\lambda)=\lambda^4-\left(\cos{\zeta_m}+\cos{\zeta_n}+\cos{\zeta_n}\cos{\theta}-\cos{\zeta_m}\cos{\theta}\right) \lambda^3 +\left(\cos{\zeta_m}+\cos{\zeta_n}+\cos{\zeta_n}\cos{\theta}-\cos{\zeta_m}\cos{\theta}\right) \lambda-1.$$ Solving $\chi_{\mathrm{U'''}_{n,m}}(\lambda)=0$ we get the required eigenvalues. $\hfill{\square}$

\begin{figure}[H] 
\centering
\subfigure[ $\overline{P}_{\infty}(\psi_R),\overline{P}_{\infty}(\psi_L)$]{\includegraphics[height=5.7 cm,width=6.5 cm]{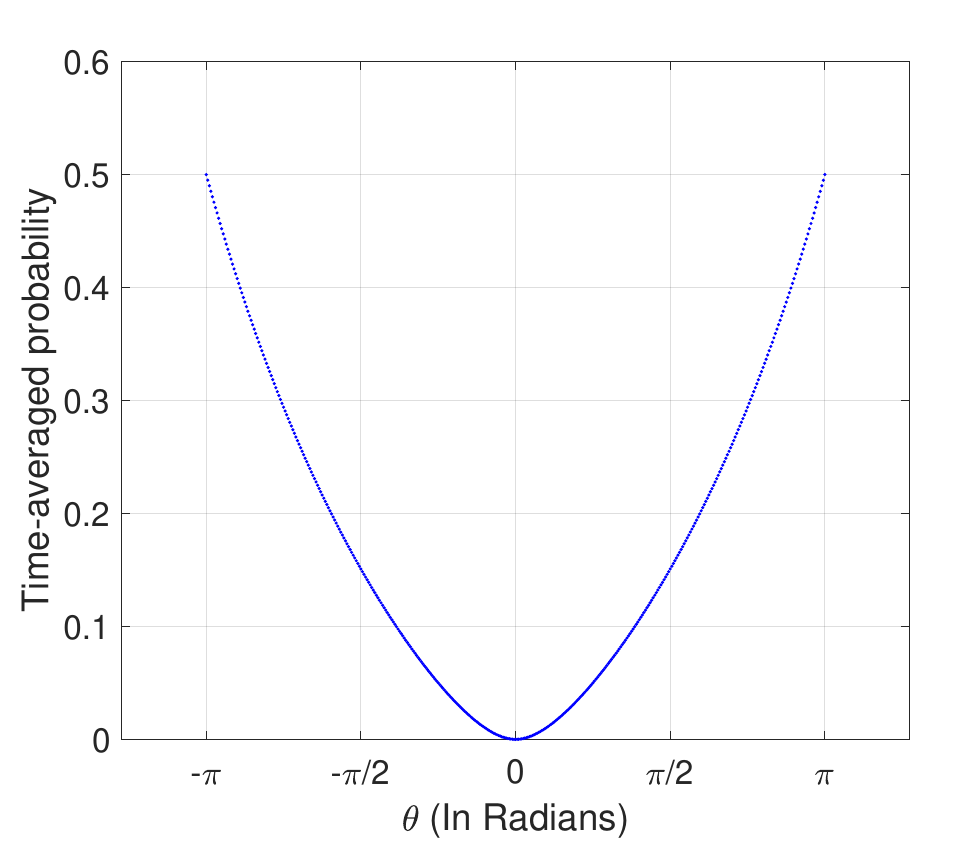}} 
\hspace{0.3cm}
\subfigure[ $\overline{P}_{\infty}(\psi_U),\overline{P}_{\infty}(\psi_D)$]{\includegraphics[height=6 cm,width=6.5 cm]{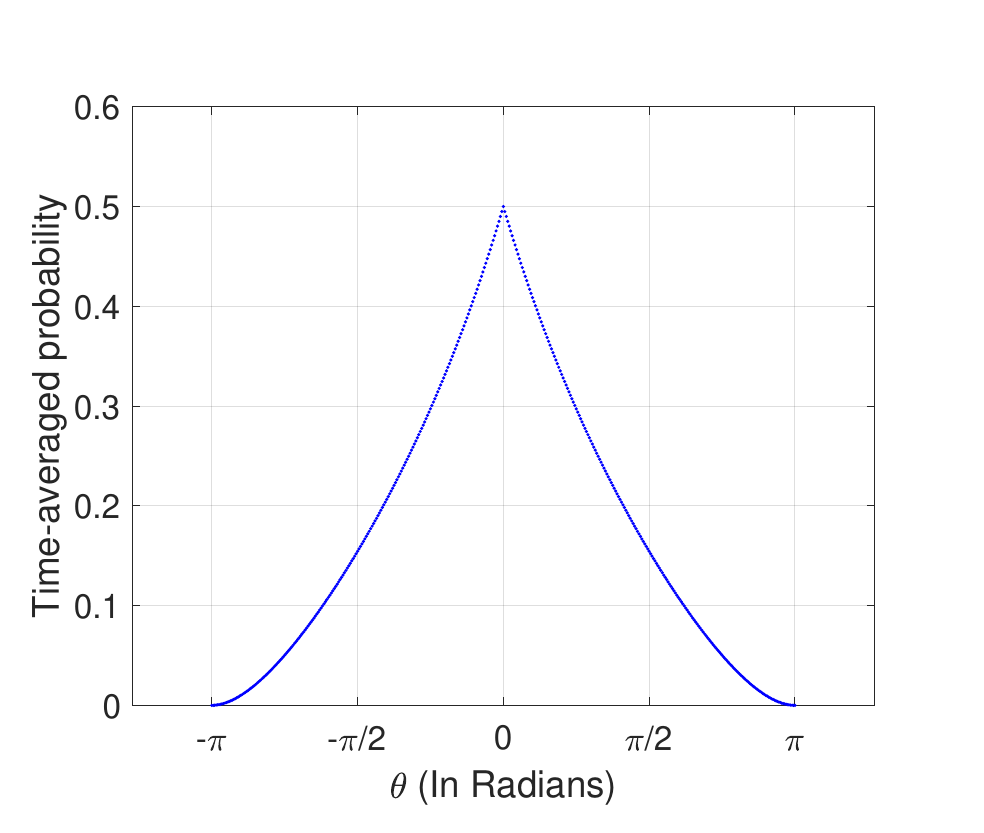}}
\caption{Numerical values of $\overline{P}_\infty ((0,0);\psi_S(0)), S\in\{R, L, U, D\}$ when the coins belong to $W_\theta$ for $400$  equidistant values of $\theta$ in $[-\pi, \pi].$}\label{fig5} 
\end{figure}

Adapting similar procedures as in the above subsections, the values of total time-averaged probabilities $\overline{P}_\infty ((0,0);\psi_S(0)), S\in\{R, L, U, D\}$ can be determined numerically for canonical initial coin states for infinite lattice. In Figure \ref{fig5}, we plot  $\overline{P}_\infty ((0,0);\psi_S(0))$ for different values of $\theta$ obtained by discretizing the interval $[-\pi, \pi]$ into $400$ equidistant points. Then observe that the total probabilities are symmetric about the vertical axis through $\theta=0,$ where it becomes zero for initial coin state $\ket{\psi_S(0)},$ $S\in\{R, L\}.$ Indeed, $\overline{P}_\infty ((0,0);\psi_S(0))$ increases and decreases as $|\theta|$ increases for  $S\in\{R,L\}$ and $S\in\{U,D\},$ respectively. Besides, observe that the total probabilities $\overline{P}_\infty ((0,0);\psi_S(0)),$ $S\in\{U,D\}$ approach to zero for several coins when  $|\theta|$ equals to $\pi.$ 

 In following proposition we determine the initial coin states for which the proposed quantum walks give the total time-averaged probability zero when $C\in W_\theta.$
 
 
 \begin{proposition}\label{pro:ini st W per}
 The walk with $C\in W_\theta,\theta\in [-\pi,\pi],$ does not show localization at the initial position $(0,0)\in \mathbb{Z}\times \mathbb{Z}$ if and only if the initial coin state $\ket{\psi_{0,0}(0)}\in I^{(w)}_0$ where
 $$I^{(w)}_0=\left\{[a,b,c,d]^T\in \mathbb{C}^4|a=b=\alpha(1+\cos \theta),c=-d=\alpha\sin \theta,\alpha\in \mathbb{C},|\alpha|^2=\frac{1}{4 (1+\cos \theta)}\right\}.$$ if $\theta\notin\{0,\pm \pi\},$ $I^{(w)}_0=\{[\alpha,\beta,0,0]^T|\alpha,\beta\in \mathbb{C},|\alpha|^2+|\beta|^2=1\}$ if $\theta=0,$ and  $I^{(w)}_0=\{[0,0,\alpha,\beta]^T|\alpha,\beta\in \mathbb{C},|\alpha|^2+|\beta|^2=1\},$ if $\theta=\pm \pi.$
\end{proposition}
\pf The proof follows similarly as Proposition \ref{pro:ini st Y}, by using Lemma \ref{lem:eig W}. $\hfill{\square}$
 

Note that, for the walk with $C\in W_\theta$ we obtain $\overline{P}_\infty ((0,0);\psi_S(0))=0,$ for $S\in\{R, L\}$ when $\theta =0$ and  $S\in\{U, D\}$ when $\theta =\pm \pi$ by Proposition \ref{pro:ini st W per}. These results can be observed in Figure \ref{fig5}. 


\begin{remark} \label{remark:end}
From Remark \ref{remark:ev} and \ref{remark:infi N}, it follows that localization property of the proposed quantum walks on infinite lattice depends on the constant eigenvalues of the evolution operator, and if initial state belongs to the orthogonal complement of the eigenspaces corresponding to the constant eigenvalues then the walks do not localize at the initial position, which is also established in \cite{tate2019eigenvalues} for periodic evolution operator and speculated in several other articles. The framework of investigating the asymptotic behavior of quantum walks on a finite lattice and then extending it to infinite lattice helps deriving a computable formula for the total time-averaged probability of finding the walker at its initial position for infinite lattice. The technique of approximating the probability amplitudes of the quantum state of the walker through an integral formula enables us to calculate the total time-averaged probability in terms of the coin parameter $\theta.$ This reveals how the total time-averaged probability depends on the coin parameter that can be used for specific applications. 
\end{remark}

\noindent{\bf Conclusion.} In this paper, we study discrete-time four-state quantum walks on two-dimension lattices. The coins are considered as one-parameter orthogonal matrices that are also permutative, a property of the Grover matrix. We focus on four sets of parametric coins denoted as $X_\theta, Y_\theta, Z_\theta$ and $W_\theta$ such that the Grover matrix belongs to the first three classes for some value of $\theta.$ We perform a thorough analysis of localization phenomena of these walks. We show that walks with coin operators from $X_\theta,Y_\theta$ and $Z_\theta$ localize at the initial position of the walker, which is considered as the origin of the two-dimensional lattices, for canonical initial coin states. However, this does not hold true for all coins in $W_\theta.$ We also show that for coins in $X_\theta,\theta\neq \pi/2$ the walks localize for any initial coin state, whereas we provide a complete characterization of initial coin states for walks with coins from $Y_\theta, Z_\theta, W_\theta$ for which the corresponding walks do not localize at the initial position. \\

 


\noindent{\bf Acknowledgement.} Amrita Mandal thanks Council for Scientific and Industrial Research (CSIR), India for financial support in the form of a junior/senior research fellowship. Rohit Sarma Sarkar acknowledges support through Prime Minister Research Fellowship (PMRF), Government of India.

\end{document}